\documentclass[conference]{IEEEtran}
\usepackage{amsmath}
\usepackage{caption}
\usepackage{subcaption}
\usepackage{framed}
\usepackage{cancel}

\usepackage{rotating}
\usepackage{tikz}
\usepackage{pgfplots}

\usepackage{amssymb}
\usepackage{hyperref}
\usepackage{longtable}

\allowdisplaybreaks

\ifCLASSINFOpdf
\else
\fi

\begin{document}
%

\title{Learning Directed-Acyclic-Graphs from Large-Scale Genomics Data}

\author{\IEEEauthorblockN{Fabio~Nikolay\IEEEauthorrefmark{2},
Marius~Pesavento\IEEEauthorrefmark{2},
George~Kritikos\IEEEauthorrefmark{3},
Nassos~Typas\IEEEauthorrefmark{3}}
\IEEEauthorblockA{\IEEEauthorrefmark{2}Communication Systems Group, Technische Universit{\"a}t Darmstadt, Germany}
\IEEEauthorblockA{\IEEEauthorrefmark{3}European Molecular Biology Laboratory (EMBL), Heidelberg, Germany}
}




\maketitle

\begin{abstract}
In this paper we consider the problem of learning the genetic-interaction-map, i.e., the topology of a directed acyclic graph (DAG) of genetic interactions from noisy double knockout (DK) data. Based on a set of well established biological interaction models we detect and classify the interactions between genes.
We propose a novel linear integer optimization program called the Genetic-Interactions-Detector (GENIE) to identify the complex biological dependencies among genes and to compute the DAG topology that matches the DK measurements best.
Furthermore, we extend the GENIE-program by incorporating genetic-interactions-profile (GI-profile) data to further enhance the detection performance. 
In addition, we propose a sequential scalability technique for large sets of genes under study, in order to provide statistically stressable results for real measurement data.  
Finally, we show via numeric simulations that the GENIE-program as well as the GI-profile data extended GENIE (GI-GENIE)-program clearly outperform the conventional techniques and present real data results for our proposed sequential scalability technique.
\end{abstract}


%
\IEEEpeerreviewmaketitle

\section{Introduction}
\IEEEPARstart{G}{enetic} interaction analysis aims at uncovering the interactions among a set of genes with respect to a specified cell function of a biological system, e.g., the fitness of a specific bacteria colony. The interactions among the genes under study can be characterized by a directed-acyclic-graph (DAG)\cite{dag_paper} where the hierarchical relationship among two genes of a DAG describes their hierarchical interaction type\cite{Battle}. However DAGs cannot be observed directly but only the specified cell function under study which yields observable phenotypes.
The term phenotype generally describes the specific manifestation of a biological attribute of an organism which can be observed, e.g., for bacteria a common biological attribute is the growth measured in colony size, where a specific size of the bacteria colony is a phenotype of this biological attribute.
The role of the studied genes in the cell machinery, the hierarchical interaction types of the genes, as well as the DAG, which describes the latter ones, can only be learned by means of knock-out experiments where a gene or a set of genes is functionally switched off and the phenotype is observed.
Traditionally, only single-knock-out (SK) experiments have been conducted but those mainly provide evidence on the importance of a single gene for the investigated cell process and do not convey much information about the interaction among the genes under study.    
\\
Recently, with the technological advances in micro arrays and the development of the synthetic-genetic-array technologies \cite{SGA}
new approaches have been taken that are based on large scale knock-out experiments of pairs of genes. Such double knock-out (DK) experiments are much more powerful for exploring genetic interactions since a DK phenotype of an arbitrary pair of genes generally differs considerably from the superposition of the corresponding SK phenotypes of this pair of genes. 
According to \cite{Battle}, the gene pairs can be classified into one out of five hierarchical relationship classes based on their SK and DK phenotypes.   
Further, based on the hierarchical relationship classes the DAG underlying the observed SK and DK phenotypes can be inferred which directly reflects the genetic interactions among the genes.   
In order to detect the DAG underlying the SK and DK phenotypes a variety of statistical methods based on scoring the measurements or on thresholding the genetic-interaction (GI)-profile data, which is commonly based on Pearson correlation of the SK and DK phenotypes, 
 e.g. \cite{IEEEhowto:Snijder}-\cite{Brock} respectively, have been developed. However, methods as presented in \cite{IEEEhowto:Snijder}-\cite{Brock} have three considerable disadvantages:~\emph{D1}) they show poor performance in detecting the DAG underlying the observed SK and DK phenotypes~\emph{D2}) they have no ability to combine different types of side information, e.g., GI-profile data with SK and DK phenotypes, to enhance the detection quality~\emph{D3}) they cannot make use of prior knowledge in order to enhance the DAG detection quality. 
Especially the ability to overcome the disadvantage in \emph{D2}) will become more important in the future, since there is a constantly increasing amount of different data types, e.g., SK and DK phenotypes, Pearson correlation based GI-profile data, other types of GI-profile data, freely available.    
Furthermore, the ability to overcome the deficit in \emph{D3}), i.e., to incorporate a priori knowledge about existing results in genomics research into the DAG detection procedure, is also of great significance, since existing functional relationships among genes are increasingly better understood based on a variety of studies that constantly extend the knowledge on the cell machinery and molecular biology. 
Although exhibiting the above mentioned disadvantages \emph{D1}) to \emph{D3}), methods as those presented in \cite{IEEEhowto:Snijder}-\cite{Brock} are the most commonly used methods to detect the DAG underlying the measured SK and DK data.
\\
Therefore, we propose the Genetic-Interactions-Detector (GENIE)-program, that is an approach based on the biological system model of \cite{Battle} with which it is possible to overcome the above mentioned shortcomings of the most popular methods as those reported in \cite{IEEEhowto:Snijder}-\cite{Brock}. 
Since the hierarchical relationship classes are mutually dependent, classifying each pair of genes to a specific hierarchical relationship class corresponds to a multi-hypotheses test.
Thus, we formulate this multi-hypotheses test as a linear integer optimization program, \cite{LIP1}-\cite{LIP6} in order to find the set of hierarchical relationship classes, best matching the observed SK and DK phenotypes.   
Based on the detected set of hierarchical relationship classes, the set of edges of the DAG which reflects the interactions among the genes can be computed.
Furthermore, we propose the GI-GENIE-program where we advance the proposed GENIE-program by incorporating GI-profile data, e.g. GI-profile data based on Pearson correlation of the observed SK and DK phenotypes, into the DAG detection procedure. 
Due to incomplete knowledge about the true dependencies among the very most sets of genes, i.e., the true DAG of a set of genes with respect to a specific cell function is unknown or only partially known for almost all sets of genes irrespectively of the cell function under study, there is a strong interest in the genomics research community in statistically reliable statements about the topology of the DAGs underlying large sets of genes, i.e., for the empirical probability of a pair of genes to interact with each other.
Towards this aim we propose a sequential technique based on the GENIE/GI-GENIE algorithms that yields statistically stressable 
statements about the interactions among genes from a large set of genes under study.
\\ 
This paper is organized as follows. We first summarize the biological system model of \cite{Battle} in Section~2, then we present in Section~3 the GENIE-program for detecting the set of hierarchical relationship classes, that represents a valid DAG and matches the DK measurements best.
In Section~4 we extend the GENIE-program with GI-profile data (GI-GENIE).
In Section~5, we present our scalability approach in order to obtain statistically stressable results for large sets of genes.
Following Section~5, we present results for simulated data which demonstrate the performance of the GENIE and the GI-GENIE methods in Section~6. Furthermore, in Section~6 we display real data results for the scalability approach described in Section~5.   
Finally, we summarize in Section~7 the key parts of this paper and give a brief outlook on future work.                

\section{System Model}
In this section we provide a mathematical description of a DAG as well as its biological implications. Furthermore, we introduce the common biological terms and provide a compact description of the genetic interaction model of \cite{Battle} including simple explanations on how to read and interpret a DAG of a genetic interaction map. \\ 
According to \cite{Graph} a graph $\mathcal{A} = \big( \text{V}(\mathcal{A}), \text{E}(\mathcal{A}) \big) $ is well defined by a set of nodes $\text{V}(\mathcal{A}) = \left\{ \text{a}_1,\text{a}_2,...,\text{a}_A \right\} $ and a set of edges $\text{E}(\mathcal{A})= \big\{ \left\{ \text{a}_1, \text{a}_A,\right\},\left\{ \text{a}_2, \text{a}_A,\right\}, ..., \left\{ \text{a}_A, \text{a}_1,\right\} \big\} $ where $\left\{ \text{a}_i, \text{a}_j,\right\}$ for $\text{a}_i,\text{a}_j \in \text{V}(\mathcal{A})$ denotes a directed edge from $\text{a}_i$ to $\text{a}_j$ and cardinality $\left| \text{V}(\mathcal{A}) \right| = A$ denotes the number of elements of set $\text{V}(\mathcal{A})$. 
The operators $\text{V}(\cdot)$ and $\text{E}(\cdot)$ applied to graph $\mathcal{A}$ yield the set of nodes $\text{V}(\mathcal{A})$ and the set of edges $\text{E}(\mathcal{A})$ respectively. We mostly address sets $\text{V}(\mathcal{A})$ and $\text{E}(\mathcal{A})$ by $\mathcal{G}_{\mathcal{A}}$ and $\mathcal{E}_{\mathcal{A}}$ respectively for the sake of notational convenience, i.e., $\mathcal{A} = \big( \mathcal{G}_{\mathcal{A}}, \mathcal{E}_{\mathcal{A}} \big) $.
Assume that there is a path P from node $a_i\in \mathcal{G}_{\mathcal{A}}$ to node $a_j \in \mathcal{G}_{\mathcal{A}}$ in graph $\mathcal{A}$, i.e., a directed connection from node $a_i\in \mathcal{G}_{\mathcal{A}}$ to node $a_j\in \mathcal{G}_{\mathcal{A}}$. Then path P is described by the concatenation of nodes being passed through on the way from node $a_i\in \mathcal{G}_{\mathcal{A}}$ to node $a_j\in \mathcal{G}_{\mathcal{A}}$, i.e., $\text{P} = a_i...a_j$ and $\text{V}(\text{P})= \left\{ a_i, ..., a_j\right\}$ denotes the set of nodes of path P \cite{Graph}.  
\\
The functional dependencies among a set of genes
$\mathcal{G} = \left\{ g_1, ..., g_G \right\}$, with $G = \left| \mathcal{G} \right|$ elements, for a given cell process and specie can be characterized by a genetic-interaction-map (GI-map,\cite{GI_map1}-\cite{GI_map5}) which is essentially a DAG with a common root node, i.e., the reporter level $R$,\cite{fabio_eusipco}. In particular, an arbitrary DAG $\mathcal{D}$ can be described as a graph $\mathcal{D} = \left(\mathcal{G}_\mathcal{D},\mathcal{E}_\mathcal{D} \right)$ with the set of nodes $\mathcal{G}_\mathcal{D} = \left\{ \mathcal{G} \cup R\right\}$ and the set of directed edges $\mathcal{E}_\mathcal{D} = \Big\{ \left\{ g_i, g_j \right\},..., \left\{ g_j, g_l  \right\}   \Big\} $. As the genetic interactions can only be observed through the reporter, all edges are always orientated in such a way that each path parting from any arbitrary gene $g_i \in \mathcal{G}$ always terminates in the root node $R$ and any gene appears on the path at most once, i.e., there exist no cycles in the graph. Hence, the DAG $\mathcal{D}$ is always connected via its root node $R$.  
For the sake of notational convenience, in most cases we write gene $i$ when addressing gene $g_i$,\cite{fabio_eusipco}.
The reporter node $R$ is an artificial node, i.e., not a gene, in the concept of a DAG and represents the measured phenotype of the specific cell process under study. 
\begin{figure}

\begin{center}
\begin{tikzpicture}
\begin{scope}[scale=.75]
\node[draw,circle, fill=gray!80,inner sep=0pt,minimum size=10pt ] (n1) at (0,0){R};
\node[draw,circle,fill=gray!20,inner sep=0pt,minimum size=10pt ] (n2) at (0,1){$n_0$};
\node[draw,circle,fill=gray!20,inner sep=0pt,minimum size=10pt] (n3) at (0,2){};
\node[draw,circle,fill=gray!20,inner sep=0pt,minimum size=10pt] (n4) at (0,3){$l_0$};
\node[draw,circle,fill=gray!20,inner sep=0pt,minimum size=10pt] (n5) at (-1,4){$g_0$};
\node[draw,circle,fill=gray!20,inner sep=0pt,minimum size=10pt] (n6) at (-1,5){};
\node[draw,circle,fill=gray!20,inner sep=0pt,minimum size=10pt] (n7) at (-1,6){$j_0$};
\node[draw,circle,fill=gray!20,inner sep=0pt,minimum size=10pt] (n8) at (-1,7){$i_0$};

\node[draw,circle,fill=gray!20,inner sep=0pt,minimum size=10pt] (n9) at (1,4){$t_0$};
\node[draw,circle,fill=gray!20,inner sep=0pt,minimum size=10pt](n10) at (1,5){};
\node[draw,circle,fill=gray!20,inner sep=0pt,minimum size=10pt] (n11) at (1,6){};

\node[draw,circle,fill=gray!20,inner sep=0pt,minimum size=10pt] (n12) at (3,4){};
\node[draw,circle,fill=gray!20,inner sep=0pt,minimum size=10pt] (n13) at (3,5){};
\node[draw,circle,fill=gray!20,inner sep=0pt,minimum size=10pt] (n14) at (3,6){};

\path[-latex,] (n2) edge (n1);
\path[-latex,] (n3) edge (n2);
\path[-latex,] (n4) edge (n3);
\path[-latex,] (n5) edge (n4);
\path[-latex,] (n6) edge (n5);
\path[-latex,] (n7) edge (n6);
\path[-latex,] (n8) edge (n7);

\path[-latex,] (n9) edge (n4);
\path[-latex,] (n10) edge (n9);
\path[-latex,] (n11) edge (n10);

\path[-latex,] (n12) edge (n3);
\path[-latex,] (n13) edge (n12);
\path[-latex,] (n14) edge (n13);

\path[-latex,out=190,in=130] (n4) edge (n1);
\path[-latex,out=270,in=45] (n12) edge (n1);

\end{scope}

\end{tikzpicture}
\end{center}

\label{fig:example_DAG}
\caption{DAG $\mathcal{D}_0$ of 13 genes and root node R}

\end{figure}
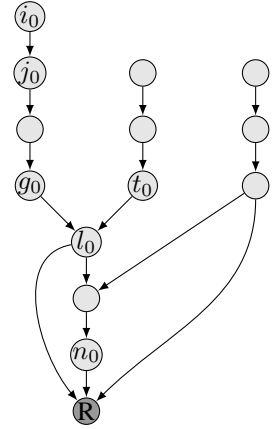
\\
To provide a better understanding of the information encoded in a DAG we state a simple example, which is similar to the one in \cite{fabio_eusipco}, based on DAG $\mathcal{D}_0$ displayed in Fig.~1. In $\mathcal{D}_0$ there exists an direct edge from gene $i_0$ to gene $j_0$, i.e. $\left\{ i_0, j_0 \right\} \in \mathcal{E}_{\mathcal{D}_0}$, which indicates that the activity of gene $i_0$ controls the activity of gene $j_0$. Hence, gene $i_0$ only affects the phenotype via gene $j_0$ and not directly. We emphasize that in this model the existence of edge $\left\{ i_0, j_0 \right\}$ in the DAG only describes the hierarchical functional dependency between genes $i_0$ and $j_0$ and not the quantitative effect of gene $i_0$ on gene $j_0$.        
\\
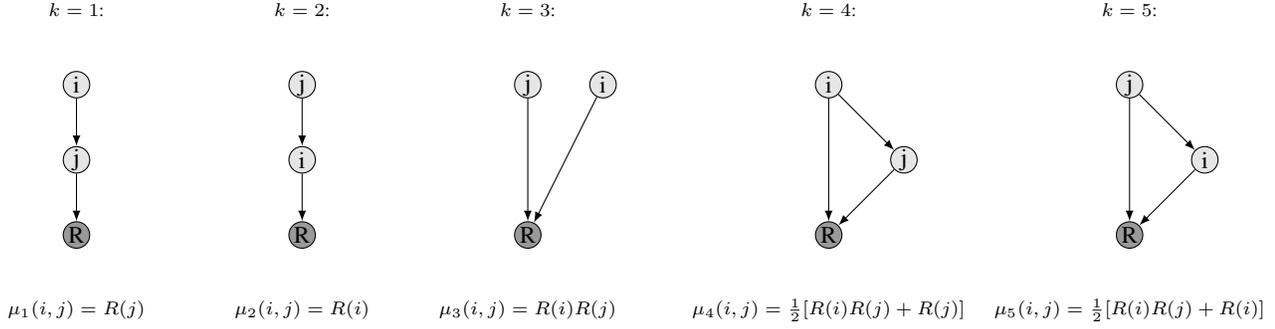
\begin{figure*}[!t]
\normalsize

\begin{tikzpicture}

\begin{scope}[scale=1]


\node (rk1) at (0,1) {\scriptsize{$k=1$:}}; 
\node[draw,circle,fill=gray!20,inner sep=0pt,minimum size=10pt] (n1) at (0,0) {\small{i}};
\node[draw,circle,fill=gray!20,inner sep=0pt,minimum size=10pt] (n2) at (0,-1) {\small{j}};
\node[draw,circle, fill=gray!80,inner sep=0pt,minimum size=10pt] (n3) at (0,-2) {\small{R}};

\path[-latex] (n1) edge (n2);
\path[-latex] (n2) edge (n3);

\node (muk1) at (0,-3) {\scriptsize{$\mu_1(i,j) = R(j)$}};

\node (rk2) at (3,1) {\scriptsize{$k=2$:}}; 
\node[draw,circle,fill=gray!20,inner sep=0pt,minimum size=10pt] (n1) at (3,0) {\small{j}};
\node[draw,circle,fill=gray!20,inner sep=0pt,minimum size=10pt] (n2) at (3,-1) {\small{i}};
\node[draw,circle,fill=gray!80,inner sep=0pt,minimum size=10pt] (n3) at (3,-2) {\small{R}};

\path[-latex] (n1) edge (n2);
\path[-latex] (n2) edge (n3);

\node (muk1) at (3,-3) {\scriptsize{$\mu_2(i,j) = R(i)$}}; 

\node (rk3) at (6,1) {\scriptsize{$k=3$:}}; 
\node[draw,circle,fill=gray!20,inner sep=0pt,minimum size=10pt] (n1) at (6,0) {\small{j}};
\node[draw,circle,fill=gray!20,inner sep=0pt,minimum size=10pt] (n2) at (7,0) {\small{i}};
\node[draw,circle,fill=gray!80,inner sep=0pt,minimum size=10pt] (n3) at (6,-2) {\small{R}};

\path[-latex] (n1) edge (n3);
\path[-latex] (n2) edge (n3);

\node (muk1) at (6,-3) {\scriptsize{$\mu_3(i,j) = R(i)R(j)$}};

\node (rk4) at (10,1) {\scriptsize{$k=4$:}}; 
\node[draw,circle,fill=gray!20,inner sep=0pt,minimum size=10pt] (n1) at (10,0) {\small{i}};
\node[draw,circle,fill=gray!20,inner sep=0pt,minimum size=10pt] (n2) at (11,-1) {\small{j}};
\node[draw,circle,fill=gray!80,inner sep=0pt,minimum size=10pt] (n3) at (10,-2) {\small{R}};

\path[-latex] (n1) edge (n3);
\path[-latex] (n1) edge (n2);
\path[-latex] (n2) edge (n3);

\node (muk1) at (10,-3) {\scriptsize{$\mu_4(i,j) = \frac{1}{2}[R(i)R(j) + R(j)]$}};

\node (rk5) at (14,1) {\scriptsize{$k=5$:}}; 
\node[draw,circle,fill=gray!20,inner sep=0pt,minimum size=10pt] (n1) at (14,0) {\small{j}};
\node[draw,circle,fill=gray!20,inner sep=0pt,minimum size=10pt] (n2) at (15,-1) {\small{i}};
\node[draw,circle,fill=gray!80,inner sep=0pt,minimum size=10pt] (n3) at (14,-2) {\small{R}};

\path[-latex] (n1) edge (n3);
\path[-latex] (n1) edge (n2);
\path[-latex] (n2) edge (n3);
\node (muk1) at (14,-3) {\scriptsize{$\mu_5(i,j) = \frac{1}{2}[R(i)R(j) + R(i)]$}}; 

\end{scope}
\end{tikzpicture}

\hrulefill

\label{fig:relationshipsBattle}
\caption{Possible hierarchical relationship classes between two arbitrary genes $i,j$ of DAG $\mathcal{D}$ according to \cite{Battle}}

\end{figure*} 
Denote $R(i) \in \mathbb{R}$ as the phenotype for a single gene $i \in \mathcal{G}$ functionally switched off.
In the same fashion we define the phenotype for the DK of genes $i,j \in \mathcal{G}$$:j>i$ as $R(i,j) \in \mathbb{R}$. 
Based on the DK phenotypes the GI-profile similarity of genes $i,j \in \mathcal{G}$$:j>i$ is computed as the Pearson correlation between all DK phenotypes which involve gene $i$ and all DK phenotypes which involve gene $j$, i.e., the GI-profile similarity $\rho(i,j)$ is given by 
\begin{align}
\rho(i,j) = \frac{ \sum\limits_{l \in \mathcal{G} \setminus i,j} \left( R(i,l) - \bar{R}(i)  \right) \left( R(j,l) - \bar{R}(j)  \right)  }{ 
\sqrt{ \sum\limits_{l \in \mathcal{G} \setminus i,j} \left(  R(i,l) -\bar{R}(i)   \right)^2 \sum\limits_{l \in \mathcal{G} \setminus i,j} \left(  R(j,l) -\bar{R}(j)   \right)^2  } }
\label{GI_data} 
\end{align}  
with $\bar{R}(i) = \frac{1}{G-2} \sum\limits_{l \in \mathcal{G} \setminus i,j} R(i,l)$, $\bar{R}(j) = \frac{1}{G-2} \sum\limits_{l \in \mathcal{G} \setminus i,j} R(j,l)$ and $R(i,l) = R(l,i) \ \forall i,l$$ \in \mathcal{G}$$:l>i$. 
We remark that the GI-profile similarity data $\rho(i,j)$ does not have to be computed according to Eq.~(\ref{GI_data}) necessarily. It can also be extracted from a data base where a priori knowledge about the set of genes under study, i.e., $\mathcal{G}$, is stored.  
 In genomics research it is a common assumption that if there is an edge between two genes $i,j \in \mathcal{G}$$:j>i$ in DAG $\mathcal{D}$, i.e., there is an interaction between genes $i,j \in \mathcal{G}$$:j>i$ in DAG $\mathcal{D}$, then the GI-profile similarity $\rho(i,j)$ is very likely to be high.  
Furthermore according to \cite{Battle} we assume that each pair of genes $i,j \in \mathcal{G}:$$j>i$ belongs to exactly one out of five hierarchical relationship classes that are characterized in Fig.~2.
The hierarchical relationship classes $k \in \mathcal{K} =\left\{1,...,5 \right\} $ are defined according to the model $\mu_k(i,j)$ in which the single knock-out phenotypes $R(i)$ and $R(j)$ are related with the DK phenotype $R(i,j)$. 
If the gene pair $i,j \in \mathcal{G}$$:j>i$ belongs to the hierarchical relationship class $k$ then the observed DK phenotype $R(i,j)$ is described by the model $\mu_k(i,j)$ provided in Fig.~2. 
We remark that the five hierarchical dependency graphs in Fig.~2 do not reflect the absolute adjacency relations, but the hierarchical relations between genes $i,j$ in DAG $\mathcal{D}$. Hence, given that two genes $i,j$ of DAG $\mathcal{D}$ are in class $k$, we cannot conclude that genes $i,j$ are directly arranged in DAG $\mathcal{D}$ as displayed by the depiction of class $k$ in Fig.~2. This follows from the fact that the description of the hierarchical relationship classes provided in Fig.~2 only contains relative topology information about two genes $i,j$ in DAG $\mathcal{D}$. 
\\
In the following and in addition to \cite{Battle}, we provide, for clarity of presentation, a formal description of the hierarchical relationship classes depicted in Fig.~2 using a graph theoretical representation.
Assume that there are $I$ paths $\text{P}_{i,\tau}, \text{ for } \tau \in \left\{1,...,I\right\}$, from gene $i$ to the reporter node $R$ in DAG $\mathcal{D}$ and the set $\mathcal{P}_{i}$ containing all such paths is defined as    
$\mathcal{P}_{i} = \left\{ \text{P}_{i,1},...,\text{P}_{i,I}  \right\}$. Furthermore, set $\mathcal{P}_{j} = \left\{ \text{P}_{j,1},...,\text{P}_{j,J}  \right\}$ contains all $J$ paths from gene $j$ to the reporter node $R$ in DAG $\mathcal{D}$. Given gene pair $i,j \in \mathcal{G}:j>i$ in DAG $\mathcal{D}$, then pair $i,j$ belongs to the hierarchical relationship class $k \in \mathcal{K}$ if and only if condition $\text{C}_k$ as defined below is satisfied:
\begin{subequations}
\begin{align}
\text{C}_1: \nonumber \\
\begin{split} \hspace{8em}
 \forall \ \text{P}_{i,\tau} \in \mathcal{P}_i : j \in \text{V}(\text{P}_{i,\tau}) \label{cond_1} 
\end{split} \\
\text{C}_2: \nonumber \\
\begin{split} \hspace{8em}
 \forall \ \text{P}_{j,\tau} \in \mathcal{P}_j : i \in \text{V}(\text{P}_{j,\tau}) \label{cond_2}
\end{split} \\
\text{C}_3: \nonumber \\ 
\begin{split} \hspace{6em}
 \Big( \forall \ \text{P}_{i,\tau} \in \mathcal{P}_i : j \notin \text{V}(\text{P}_{i,\tau}) \Big) \bigwedge \\
 \Big( \forall \ \text{P}_{j,\tilde{\tau}} \in \mathcal{P}_j : i \notin \text{V}(\text{P}_{j,\tilde{\tau}}) \Big) \label{cond_3}
\end{split} \\
\text{C}_4: \nonumber \\
\begin{split}
 \Big( \exists \ \text{P}_{i,\tau} \in \mathcal{P}_i: j \notin \text{V}(\text{P}_{i,\tau})   \Big) \bigwedge \\
 \Big( \exists \ \text{P}_{i,\tau} \in \mathcal{P}_i, \text{P}_{j,\tilde{\tau}} \in \mathcal{P}_j  : \text{V}(\text{P}_{j,\tilde{\tau}} ) \subset  \text{V}(\text{P}_{i,\tau} )  \Big) \label{cond_4}
\end{split} \\
\text{C}_5: \nonumber \\
\begin{split}
 \Big(\exists \ \text{P}_{j,\tilde{\tau}} \in \mathcal{P}_j: i \notin \text{V}(\text{P}_{j,\tilde{\tau}}) \Big) \bigwedge \\ 
\Big( \exists \ \text{P}_{i,\tau} \in \mathcal{P}_i, \text{P}_{j,\tilde{\tau}} \in \mathcal{P}_j  : \text{V}(\text{P}_{i,\tau} ) \subset \text{V}(\text{P}_{j,\tilde{\tau}} )    \Big) \label{cond_5}
\end{split} 
\end{align}
\end{subequations}     
As stated in condition $\text{C}_1$ in (\ref{cond_1}), two genes $i,j \in \mathcal{G}$$:j>i$ in DAG $\mathcal{D}$ belong to the hierarchical relationship class $k=1$, if all paths from gene $i$ to the reporter node $R$ pass through gene $j$. Hence, gene $j$ is always an element of the set of nodes of each path $\text{P}_{i,\tau} \in \mathcal{P}_i$ from gene $i$ to the reporter node $R$, i.e., $j \in \text{V}( \text{P}_{i,\tau}  )$ for all paths $\text{P}_{i,\tau}$ from gene $i$ to the reporter node $R$.      
With the same line of argument as used in (\ref{cond_1}), two genes $i,j \in \mathcal{G}$$:j>i$ in DAG $\mathcal{D}$ belong to the hierarchical relationship class $k=2$ if condition $\text{C}_2$ in (\ref{cond_2}) is satisfied.
Two genes $i,j \in \mathcal{G}$$:j>i$ in DAG $\mathcal{D}$ belong to the hierarchical relationship class $k=3$ and are considered to be independent from each other if condition $\text{C}_3$ in (\ref{cond_3}) is satisfied which states that there is no path $\text{P}_{i,\tau}$ from gene $i$ to the reporter node $R$ that passes through gene $j$ as well as there is no path $\text{P}_{j,\tilde{\tau}}$ from gene $j$ to the reporter node $R$ that passes through gene $i$.   
As stated in (\ref{cond_4}), two genes $i,j \in \mathcal{G}$$:j>i$ in DAG $\mathcal{D}$ belong to the hierarchical relationship class $k=4$ if there is at least one path $\text{P}_{i,\tau}$ from gene $i$ to the reporter node $R$ which does not pass through gene $j$ as well as for all paths $\text{P}_{j,\tilde{\tau}} \in \mathcal{P}_j$ there is always a path $\text{P}_{i,\tau} \in \mathcal{P}_i$ that is a super-path of the respective $\text{P}_{j,\tilde{\tau}} \in \mathcal{P}_j$.
With the same line of argument as used in (\ref{cond_4}), two genes $i,j \in \mathcal{G}$$:j>i$ in DAG $\mathcal{D}$ belong to the hierarchical relationship class $k=5$ if condition $\text{C}_5$ in (\ref{cond_5}) is satisfied.     
\\
To illustrate this, let us consider the example DAG $\mathcal{D}_0$ of Fig.~1. All paths from gene $i_0$ to node $R$ pass through gene $j_0$, i.e., they are in a linear pathway with gene $i_0$ upwards of gene $j_0$. 
Thus the pair of genes $i_0, j_0$ belongs to class $k=1$.
Note that with the same line of argument, we conclude that also genes $i_0$ and $l_0$ belong to relationship class $k=1$.
Since all paths from gene $i_0$ to the reporter level $R$ do not pass through gene $t_0$ and all paths from gene $t_0$ to the reporter level do not pass through gene $i_0$, genes $i_0$ and $t_0$ belong to the hierarchical relationship class $k=3$ as given in Fig.~2, which states that genes $i_0$ and $t_0$ are independent of each other and the DK phenotype amounts to $R(i_0,t_0) = \mu_3(i_0,t_0)$.  
Finally, let us investigate the hierarchical relation between genes $t_0$ and $n_0$ in DAG $\mathcal{D}_0$. Obviously, gene $t_0$ has (at least) one path to node $R$ which does not pass through gene $n_0$, i.e., genes only having paths to $R$ that do not pass through gene $n_0$ do not affect the activity of gene $n_0$. Since there is (at least) one other path from gene $t_0$ to $R$ passing through gene $n_0$, 
we can reason that genes $t_0$ and $n_0$ belong to class $k=4$.
Generally, there are strong implications among the hierarchical relationship classes of \cite{Battle}, i.e., if some pairs belong to a specific class then this has strong implications for all other pairs.
Let us consider the case that DAG $\mathcal{D}_0$ was not known and only the hierarchical relationship classes for genes $i_0$ and $j_0$, i.e., genes $i_0$ and $j_0$ belong to class $k=1$, as well as the hierarchical relationship class for genes $i_0$ and $g_0$, i.e., genes $i_0$ and $g_0$ belong to class $k=1$ were available. 
By definition of the hierarchical dependency graphs in Fig.~2 and the assumptions, that genes $i_0$ and $j_0$ belong to class $k=1$ as well as that genes $i_0$ and $g_0$ belong to class $k=1$, we conclude that all paths from gene $i_0$ to $R$ pass through genes $j_0$ and $g_0$.
Thus, either all paths from gene $g_0$ to $R$ pass through gene $j_0$, or all paths from gene $j_0$ to $R$ pass through gene $g_0$.
Consequently, genes $j_0$ and $g_0$ either belong to the hierarchical relationship class $k=1$, or $k=2$. \\
As we have emphazised by the example above, generally if the hierarchical relationship class is known for two arbitrary genes $i,j \in \mathcal{G}:$$j>i$ as well as for another pair $i,l \in \mathcal{G}:$$l>i$, then this has strong logical implications on the hierarchical relationship classes genes $j,l \in \mathcal{G}:$$l>j$ can belong to.
Since we can interpret the classification of the pairs of genes $i,j \in \mathcal{G}: j>i$, based on their observed SK and DK phenotypes $R(i), R(j)$
and $R(i,j)$, respectively, to exactly one out of the five hierarchical relationship classes
as a coupled multi-hypotheses test, we address this problem in Section~3 by a linear integer optimization program.
The proposed linear integer optimization program identifies the most consistent set of hierarchical relationship classes, i.e., the set of hierarchical relationship classes that represents a valid DAG and matches best the DK measurements with respect to the logical coupling between the classes. 
Furthermore, in Section~4 we extend the GENIE-program proposed in Section~3 by incorporating GI-profile data in order to jointly detect the most consistent set of hierarchical relationship classes and the corresponding DAG topology.

\section{GENIE-Algorithm}
In this section, we formulate the problem of classifying the gene pairs $i,j \in \mathcal{G}:$$j>i$ into the classes of hierarchical relationships based on the observed SK and DK phenotype values as a linear integer optimization program. 
Furthermore, we translate the logical implications among the hierarchical relationship classes into constraints that ensure that 
the detected set of hierarchical relationship classes represents a valid graph. That is the detected set of hierarchical relationship classes represents a graph which is a DAG as defined in Section~2. Finally, we propose a policy to derive an estimate $\hat{\mathcal{E}}_{\mathcal{D}}$ of the true set of edges $\mathcal{E}_{\mathcal{D}}$ of DAG $\mathcal{D}$ based on the detected set of hierarchical relationship classes.
\subsection{Hierarchical Relationship Class Detection}
In order to quantify the mismatch between the measured DK phenotypes $R(i,j)$ and the phenotype model $\mu_k(i,j)$ of class $k \in \mathcal{K}$ according to Fig.~2,
under the hypothesis that the gene pairs $i,j \in \mathcal{G}:$$j>i$ belong to class $k$ given their respective SK values, we propose a simple quadratic score \cite{Battle}, \cite{fabio_eusipco},
as given in Eq.~(\ref{score}):
\begin{align}
s_k(i,j) = \big( R(i,j) - \mu_k(i,j)  \big)^2  & \quad k \in \mathcal{K}, \quad \forall i,j:\in \mathcal{G}: j>i \label{score}
\end{align} 
Let us define the following selection variables 
\begin{align}
\alpha_k(i,j) = \begin{cases} 1 & \text{ if $i,j$ are in class $k$} \\   0 & \text{ else} \end{cases} \nonumber \\
k \in \mathcal{K}, \quad \forall i,j:\in \mathcal{G}: j>i \label{relationship_indicator}
\end{align}
We remark that every DAG $\mathcal{D}$ can be represented by a set of hierarchical relationship classes which directly corresponds to a set of selection variables 	$A^{\mathcal{D}} =\bigcup\limits_{ \forall i,j \in \mathcal{G}: j>i }   \Big\{ \alpha_1^{\mathcal{D}}(i,j),..., \alpha_5^{\mathcal{D}}(i,j) \Big\} $.
The GENIE-algorithm of classifying the gene pairs $i,j \in \mathcal{G}:$$j>i$ into the set of hierarchical relationship classes, that represents
a valid DAG and matches the observed SK and DK phenotypes best
can be formulated as   
		\begin{subequations}
		\label{parent_genie}
		\begin{align} 
		\text{O}_{\text{GENIE}}: & \nonumber \\
		  \min_{\left\{ \alpha_k(i,j) \right\}} \; & \quad \sum\limits_{i=1}^{G} \sum\limits_{j=i+1}^{G}  \Big( \sum\limits_{k=1}^{\left| \mathcal{K} \right|}    s_k(i,j)\alpha_k(i,j) \Big) \label{objective_genie} \\
      \text{s. t.} \; & \quad \alpha_k(i,j) \in \left\{ 0,1 \right\} \ \forall k \in \mathcal{K}, \ \forall i,j \in \mathcal{G}: j>i \label{alpha_genie}\\
			                     & \quad \sum\limits_{k=1}^{\left| \mathcal{K} \right|} \alpha_k(i,j) = 1, \ \forall i,j \in \mathcal{G}: j>i \label{alpha_one_genie} \\
													 & \quad \mathcal{L} \Longrightarrow \text{additional topology constraints}  \label{consistency_set} 
		\end{align}
		\end{subequations}
where	$A^{\text{O}_{\text{GENIE}}} = \bigcup\limits_{ \forall i,j \in \mathcal{G}: j>i }   \Big\{ \alpha_1^{\text{O}_{\text{GENIE}}}(i,j),..., \alpha_5^{\text{O}_{\text{GENIE}}}(i,j)   \Big\} $ denotes the solution of program $\text{O}_{\text{GENIE}}$ in (\ref{parent_genie}) and the set of best matching selection variables $A^{\text{O}_{\text{GENIE}}}$ corresponds to the most consistent pattern of hierarchical relationship classes.
Problem $\text{O}_{\text{GENIE}}$ in (\ref{parent_genie}) is a linear integer program which can be solved efficiently by BB-methods \cite{BB1}-\cite{BB7}. The objective of problem $\text{O}_{\text{GENIE}}$ is to minimize the overall mismatch in classifying each gene pair $i,j \in \mathcal{G}:$$j>i$ to one out of five hierarchical relationship classes.
The constraints in (\ref{alpha_genie}) reflect the binary nature of the selection variables $\alpha_k(i,j)$, $\forall i,j \in \mathcal{G}:j>i$, $\forall k \in \mathcal{K}$, while (\ref{alpha_one_genie}) represents a multiple choice constraint to enforce that the gene pairs $i,j$ are only classified to one out of the five hierarchical relationship classes.
The set $\mathcal{L}$ in (\ref{consistency_set}) is comprised of additional constraints to ensure that the detected set of selection variables $A^{\text{O}_{\text{GENIE}}}$
always represents a valid graph, i.e., a DAG.  
In the following, we exemplarily derive topology constraints in set $\mathcal{L}$.
In order to identify the numerous logical dependencies among the selection variables $\alpha_k(i,j), k \in \mathcal{K}$ for all $i,j \in \mathcal{G}:$$j>i$ we proceed in the following way.
We first fix the assumption that genes $i,j \in \mathcal{G}:j>i$ belong to class $k=1$, i.e., $\alpha_1(i,j)=1$. Further we assume that genes $i,l \in \mathcal{G}:l>i$ belong to class $k^{'}$, i.e. $\alpha_{k^{'}}(i,l)=1$. Then we derive the set of classes $\mathcal{K}^{''}$ that genes $j,l \in \mathcal{G}:l>j$ can belong to under the assumptions made.
In the following, we have formulated the logical dependencies among the selection variables for $\alpha_1(i,j)=1$, i.e., the case that gene $i$ is linearly upstream of gene $j$, as linear integer inequalities defined in constraints (\ref{r1_1})-(\ref{r1_5}) and summarize them as set $\mathcal{L}_1$ 
\begin{subequations}
\begin{align}
\mathcal{L}_1 = \bigg\{ & \nonumber \\
& \alpha_1(j,l) +\alpha_2(j,l) \geq \alpha_1(i,j)+ \alpha_1(i,l) -1 \label{r1_1}  \\
& \alpha_2(j,l) \geq \alpha_1(i,j)+ \alpha_2(i,l) -1 \label{r1_2}  \\
& \alpha_2(j,l) + \alpha_3(j,l) + \alpha_5(j,l)  \geq \nonumber \\ 
&  \alpha_1(i,j)+ \alpha_3(i,l) -1 \label{r1_3}  \\
& \alpha_2(j,l) + \alpha_4(j,l)  \geq \alpha_1(i,j)+ \alpha_4(i,l) -1 \label{r1_4}  \\
& \alpha_5(j,l) + \alpha_2(j,l) \geq \alpha_1(i,j)+ \alpha_5(i,l) -1 \label{r1_5}  \\ 
& \bigg\}  \forall i,j,l \in \mathcal{G}_{\mathcal{D}}: l >j>i   \nonumber
\end{align} 
\end{subequations}
 where constraints (\ref{r1_1})-(\ref{r1_5}) are affine after the continuous relaxation of the selection variables $\alpha_k(i,j), \forall i,j \in \mathcal{G}:j>i$.    
To explain the origin and the functionality of the constraints in $\mathcal{L}_1$, let us further define a sub-genetic-interactions-map (SMAP) $\mathcal{S}$, \cite{fabio_eusipco}, as given in Fig.~3 according to the following definition where we adopt the graph notation of \cite{Graph}:
\\
{\bf{Definition:}}
\\ 
Given a non-empty set of edges $\mathcal{E}_{\text{in}}$ and a non-empty set of edges $\mathcal{E}_{\text{out}}$.  
Graph $\mathcal{S} = \Big(\mathcal{G}_{\mathcal{S}}, \mathcal{E}_{\mathcal{S}} \Big)$, with set of nodes $\mathcal{G}_{\mathcal{S}}$ and set of edges $\mathcal{E}_{\mathcal{S}}$, 
is a SMAP if the following conditions are fulfilled:
$i)$ the graph $\mathcal{S}$ is acyclic and directed $ii)$ there $\exists e_{\text{in}} \in \mathcal{E}_{\text{in}}, e_{\text{out}} \in \mathcal{E}_{\text{out}}$ such that each path P through graph $\mathcal{S}$ incides $\mathcal{S}$ via egde $e_{\text{in}}$ and leaves graph $\mathcal{S}$ via edge $e_{\text{out}}$.  
\begin{figure}[h!]
\centering
\begin{tikzpicture}

\begin{scope}[scale=1]


\fill[blue!10] (0,0) circle (2);
\draw[gray] (0,0) circle (2);

\node[draw,circle,fill=gray,inner sep=0pt,minimum size=10pt] (n1) at (-1,1) {};
\node[draw,circle,fill=gray,inner sep=0pt,minimum size=10pt] (n2) at (1,1) {};

\node[draw,circle,fill=gray,inner sep=0pt,minimum size=10pt] (n3) at (-1,0) {};
\node[draw,circle,fill=gray,inner sep=0pt,minimum size=10pt] (n4) at (1,0) {};

\node[draw,circle,fill=gray,inner sep=0pt,minimum size=10pt] (n5) at (0,-1) {};
\node(n_in) at (0,2.5) {};
\node(n_out) at (0,-2.5) {};

\path[-latex] (n1) edge (n3);
\path[-latex] (n3) edge (n5);

\path[-latex] (n2) edge (n4);
\path[-latex] (n4) edge (n5);

\path[-latex,blue] (n_in) edge (n1);
\path[-latex,blue] (n_in) edge (n2);
\node[blue] () at (-.7,2.1) {$e_{\text{in},1} $};
\node[blue] () at (.7,2.1) {$e_{\text{in},2} $};

\path[-latex,blue] (n5) edge (n_out);
\node[blue] (n2) at (.3,-2.5) {$e_{\text{out}} $};
\begin{scope}[scale=5]
\node () at (0,0) {$\mathcal{S}$};
\end{scope}

\end{scope}
\end{tikzpicture}

\label{fig:sub_GI_map}
\caption{Example SMAP $\mathcal{S}$}

\end{figure}
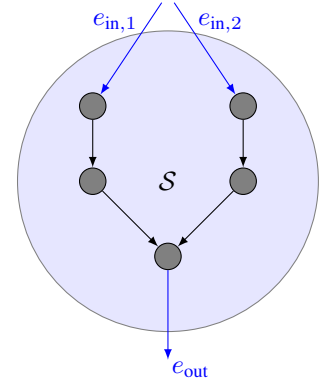  
\\
\\
DAG $\mathcal{D}_1$, as displayed in Fig.~4, consists of genes $i,j$ and SMAPs $\mathcal{S}_1$ and $\mathcal{S}_2$. Obviously, genes $i,j$ belong to class $k=1$, i.e., $\alpha_1(i,j)=1$. Furthermore, all genes $l \in \mathcal{G}_{\mathcal{D}_1} \setminus \left\{R \right\}:$$ l>j>i$ for which $\alpha_1(i,l)=1$ must be either located in SMAP $\mathcal{S}_1$ or $\mathcal{S}_2$. Thus, it follows from DAG $\mathcal{D}_1$ in Fig.~4 that the gene pair $j,l$ is either in hierarchical relationship class $k=1$ or $k=2$, i.e., $\alpha_1(j,l)=1$ or $\alpha_2(j,l)=1$. 
\\ 
This logical implication is directly reflected by constraint (\ref{r1_1}). Given $\alpha_1(i,j)=1$ and $\alpha_1(i,l)=1$ the right-hand-side (RHS) of (\ref{r1_1}) amounts to 1. In this case also the left-hand-side (LHS) of (\ref{r1_1}) becomes 1 to fulfill the inequality (\ref{r1_1}). Thus either $\alpha_1(j,l)=1$ or $\alpha_2(j,l)=1$. Reversely, assume that $\alpha_1(i,j)=1$ and $\alpha_1(i,l)=1$ does not hold, then the RHS of (\ref{r1_1}) is less than 1, i.e., 0 or -1, while the LHS of (\ref{r1_1}) 
is always greater than 0. Hence, constraint (\ref{r1_1}) is fulfilled irrespectively of the choice of $\alpha_k(j,l)$, i.e., constraint (\ref{r1_1}) enforces no logical implications.
\\
Similarly for DAG $\mathcal{D}_2$ in Fig.~4, it is obvious that genes $i,j$ belong to the hierarchical relationship class $k=1$, i.e., $\alpha_1(i,j)=1$.
All genes $l \in \mathcal{G}_{\mathcal{D}_2} \setminus \left\{R \right\}:$$ l>j>i$ which are in a linear pathway upstream of gene $i$, i.e. $\alpha_2(i,l)=1$, must be located in SMAP $\mathcal{S}_3$. Hence it directly follows from DAG $\mathcal{D}_2$ that also gene $l$ must be in a linear pathway upstream of gene $j$, i.e., $\alpha_2(j,l)=1$. This logical implication is compactly represented in constraint (\ref{r1_2}). Under the assumption that $\alpha_1(i,j)=1$ and $\alpha_2(i,l)=1$, the RHS of (\ref{r1_2}) amounts to 1 enforcing $\alpha_2(j,l)=1$, so that the LHS of (\ref{r1_2}) equals the RHS and the inequality in (\ref{r1_2}) is fulfilled. 
Reversely, assume that $\alpha_2(i,l)=0$, then the RHS of (\ref{r1_2}) is less than 1 and hence the LHS of (\ref{r1_2}) is always bigger than or equal to the RHS irrespectively of the choice of $\alpha_k(j,l)$, i.e., constraint (\ref{r1_1}) enforces no logical implications.
We can proceed in the same fashion to explain constraints (\ref{r1_3})-(\ref{r1_5}) based on DAGs $\mathcal{D}_3$ to $\mathcal{D}_5$ as given in Fig.~4 respectively. Furthermore, with the same line of argument we can derive the sets 
$\mathcal{L}_k$ for $k \in \mathcal{K} \setminus 1$ which reflect the logical implications among the selection variables under the assumptions that $\alpha_k(i,j)=1$ for $k \in \mathcal{K} \setminus 1$. However, due to space limitations we omit the derivation of the full set of logical implications at this point and refer the interested reader to \cite{supp_mat} where we will provide the full set of topology constraints $\mathcal{L}$ as well as further supplementary material.
\begin{figure*}[!t]
\normalsize

\begin{tikzpicture}

\begin{scope}[scale=1]


\node (rk1) at (0,1) {\scriptsize{$\mathcal{D}_{1}$:}}; 
\node[draw,circle,fill=gray!20,inner sep=0pt,minimum size=10pt] (n1) at (0,0) {\small{i}};
\node[draw,circle,fill=blue!30,inner sep=0pt,minimum size=20pt] (n1b) at (0,-1) {$\mathcal{S}_{1}$ };
\node[draw,circle,fill=gray!20,inner sep=0pt,minimum size=10pt] (n2) at (0,-2) {\small{j}};
\node[draw,circle,fill=blue!30,inner sep=0pt,minimum size=20pt] (n2b) at (0,-3) {$\mathcal{S}_{2}$};
\node[draw,circle, fill=gray!80,inner sep=0pt,minimum size=10pt] (n3) at (0,-4) {\small{R}};

\path[-latex] (n1) edge (n1b);
\path[-latex] (n1b) edge (n2);
\path[-latex] (n2) edge (n2b);
\path[-latex] (n2b) edge (n3);

\node (rk2) at (3,1) {\scriptsize{$\mathcal{D}_{2}$:}};
\node[draw,circle,fill=blue!30,inner sep=0pt,minimum size=20pt] (n1_1) at (3,0) {$\mathcal{S}_{3}$};
\node[draw,circle,fill=gray!20,inner sep=0pt,minimum size=10pt] (n_1_1b) at (3,-1) {\small{i}};
\node[draw,circle,fill=gray!20,inner sep=0pt,minimum size=10pt] (n_1_2) at (3,-2.75) {\small{j}};
\node[draw,circle, fill=gray!80,inner sep=0pt,minimum size=10pt] (n_1_3) at (3,-4) {\small{R}};

\path[-latex] (n1_1) edge (n_1_1b);
\path[-latex] (n_1_1b) edge (n_1_2);
\path[-latex] (n_1_2) edge (n_1_3);

\node (rk3) at (7,1) {\scriptsize{$\mathcal{D}_{3}$:}};
\node[draw,circle,fill=gray!20,inner sep=0pt,minimum size=10pt] (n3_i) at (7,0) {\small{i}};

\node[draw,circle,fill=blue!30,inner sep=0pt,minimum size=20pt] (n3_l1) at (6,-1) {$\mathcal{S}_{4}$};
\node[draw,circle,fill=blue!30,inner sep=0pt,minimum size=20pt] (n3_l2) at (8,-1) {$\mathcal{S}_{5}$};
\node[draw,circle,fill=blue!30,inner sep=0pt,minimum size=20pt] (n3_l3) at (6,-3) {$\mathcal{S}_{6}$};

\node[draw,circle,fill=gray!20,inner sep=0pt,minimum size=10pt] (n_3_j) at (7,-2.75) {\small{j}};
\node[draw,circle, fill=gray!80,inner sep=0pt,minimum size=10pt] (n_3_R) at (7,-4) {\small{R}};

\path[-latex] (n3_i) edge (n_3_j);
\path[-latex] (n_3_j) edge (n_3_R);

\path[-latex] (n3_l1) edge (n_3_j);
\path[-latex] (n3_l1) edge (n_3_R);

\path[-latex] (n3_l2) edge (n_3_j);

\path[-latex] (n3_l3) edge (n_3_R);

\node (rk4) at (11,1) {\scriptsize{$\mathcal{D}_{4}$:}};
\node[draw,circle,fill=gray!20,inner sep=0pt,minimum size=10pt] (n4_i) at (11,0) {\small{i}};

\node[draw,circle,fill=blue!30,inner sep=0pt,minimum size=20pt] (n4_l1) at (11,-1) {$\mathcal{S}_{7}$};
\node[draw,circle,fill=blue!30,inner sep=0pt,minimum size=20pt] (n4_l2) at (12,-3) {$\mathcal{S}_{8}$};

\node[draw,circle,fill=gray!20,inner sep=0pt,minimum size=10pt] (n_4_j) at (11,-2.75) {\small{j}};
\node[draw,circle, fill=gray!80,inner sep=0pt,minimum size=10pt] (n_4_R) at (11,-4) {\small{R}};

\path[-latex,out=200,in=145] (n4_i) edge (n_4_j);
\path[-latex] (n_4_j) edge (n_4_R);

\path[-latex] (n_4_j) edge (n4_l2);
\path[-latex] (n4_l2) edge (n_4_R);

\path[-latex] (n4_l1) edge (n_4_j);
\path[-latex] (n4_i) edge (n4_l1);

\node (rk5) at (15,1) {\scriptsize{$\mathcal{D}_{5}$:}};
\node[draw,circle,fill=gray!20,inner sep=0pt,minimum size=10pt] (n5_i) at (15,-2) {\small{i}};

\node[draw,circle,fill=blue!30,inner sep=0pt,minimum size=20pt] (n5_l1) at (15,0) {$\mathcal{S}_{9}$};
\node[draw,circle,fill=blue!30,inner sep=0pt,minimum size=20pt] (n5_l2) at (15,-1) {$\mathcal{S}_{10}$};

\node[draw,circle,fill=gray!20,inner sep=0pt,minimum size=10pt] (n_5_j) at (15,-3) {\small{j}};
\node[draw,circle, fill=gray!80,inner sep=0pt,minimum size=10pt] (n_5_R) at (15,-4) {\small{R}};

\path[-latex] (n5_i) edge (n_5_j);
\path[-latex] (n_5_j) edge (n_5_R);

\path[-latex] (n5_l2) edge (n5_i);
\path[-latex,out=235,in=135] (n5_l2) edge (n_5_j);

\path[-latex,out=345,in=60] (n5_l1) edge (n_5_R);
\path[-latex] (n5_l1) edge (n5_l2);

\end{scope}
\end{tikzpicture}

\hrulefill

\label{fig:relationships}
\caption{Schematically reduced DAGs $\mathcal{D}_1$ to $\mathcal{D}_5$ corresponding to Eq.~(\ref{r1_1})-Eq.~(\ref{r1_5}) respectively}

\end{figure*}
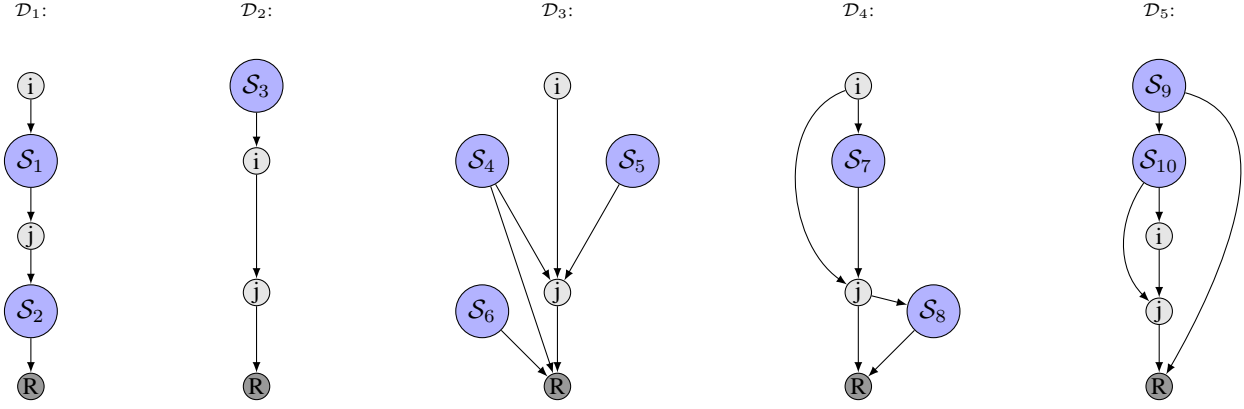 
The full set of topology constraints $\mathcal{L}$ in (\ref{consistency_set}) can be computed as 
\begin{align}
\mathcal{L} = \bigcup\limits_{k=1}^{\left| \mathcal{K} \right|} \left\{ \mathcal{L}_k \right\} \label{L}.  
\end{align}  
Finally, a considerable advantage of the presented algorithm is its ability to incorporate prior knowledge into the classification of the gene pairs $i,j \in \mathcal{G}$$:j>i$ to the most consistent hierarchical relationship classes. Assume that it is known from existing experimental results that two specific genes $i_0,j_0 \in \mathcal{G}$$:j_0>i_0$ do not interact with each other. Then we can easily incorporate this prior knowledge into program $\text{O}_{\text{GENIE}}$ in (\ref{parent_genie}) by adding Eq.~(\ref{prior_knowledge}) as defined below 
\begin{align}
\alpha_3(i_0,j_0) = 1
\label{prior_knowledge}
\end{align}   
as a topology constraint to program $\text{O}_{\text{GENIE}}$. This property is very valuable since it allows the GENIE-algorithm to take advantage of existing results in genetic interactions research to improve the reliability of the classification.

\subsection{Edge Computation}
Based on the detected set of selection variables $A^{\text{O}_{\text{GENIE}}}$ which corresponds to the most consistent pattern of hierarchical relationship classes given the observed SK and DK phenotypes, an estimate $\mathcal{E}_{\text{GENIE}}$ of the true set of edges $\mathcal{E}_{\mathcal{D}}$ of DAG $\mathcal{D}$ can be computed.  
It can be theoretically proven that the representation of an arbitrary DAG $\mathcal{D}$ by its corresponding set of hierarchical relationship classes is not unique. $\text{A}^{\mathcal{D}}$ the set of selection variables which directly corresponds to the hierarchical relationship class pattern of DAG $\mathcal{D}$ represents not only the true DAG $\mathcal{D}$, but also a set of similar DAGs which have minorly different sets of edges compared to the true DAG $\mathcal{D}$. 
Assume we are only given that $\alpha_4^{\mathcal{D}}(i,j)=1$ for two arbitrary genes $i,j \in$$\mathcal{G}:j>i$ of DAG $\mathcal{D}$, then we suffer an information loss on the number of paths from gene $i$ to the reporter node $R$ which are independent of gene $j$. Similarly, given that $\alpha_5^{\mathcal{D}}(i,j)=1$ for two arbitrary genes $i,j \in$$\mathcal{G}:j>i$ of DAG $\mathcal{D}$, we suffer an information loss on the number of paths from gene $j$ to the reporter node $R$ which are independent of gene $i$.
Hence, this information loss yields ambiguities in computing the set of edges $\mathcal{E}_{\mathcal{D}}$ of DAG $\mathcal{D}$ based on the $\text{A}^{\mathcal{D}}$.       
In order to clarify the origin of the ambiguities further, let us turn to a simple example.
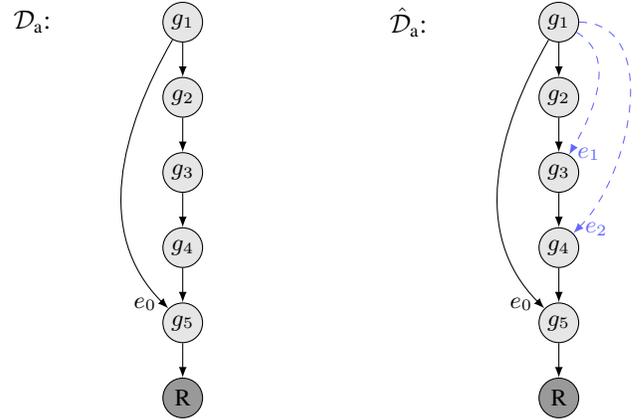
\begin{figure}[h!]
\centering
\begin{tikzpicture}

\begin{scope}[scale=1]


\node[] () at (-2,0) {$\mathcal{D}_{\text{a}}$:};
\node[draw,circle,fill=gray!20,inner sep=0pt,minimum size=15pt] (n1) at (0,0) {\small{$g_1$}};

\node[draw,circle,fill=gray!20,inner sep=0pt,minimum size=15pt] (n2) at (0,-1) {\small{$g_2$}};
\node[draw,circle,fill=gray!20,inner sep=0pt,minimum size=15pt] (n3) at (0,-2) {\small{$g_3$}};
\node[draw,circle,fill=gray!20,inner sep=0pt,minimum size=15pt] (n4) at (0,-3) {\small{$g_4$}};

\node[draw,circle,fill=gray!20,inner sep=0pt,minimum size=15pt] (n6) at (0,-4) {\small{$g_5$}};
\node[draw,circle, fill=gray!80,inner sep=0pt,minimum size=15pt] (nR) at (0,-5) {\small{R}};

\node[] () at (-.5,-3.75){\small{$e_0$}};

\path[-latex] (n1) edge (n2);
\path[-latex] (n2) edge (n3);
\path[-latex] (n3) edge (n4);
\path[-latex] (n4) edge (n6);
\path[-latex] (n6) edge (nR);

\path[-latex,out=240,in=135 ] (n1) edge (n6);


\node[] () at (3,0) {$\hat{\mathcal{D}}_{\text{a}}$:};
\node[draw,circle,fill=gray!20,inner sep=0pt,minimum size=15pt] (n1b) at (5,0) {\small{$g_1$}};

\node[draw,circle,fill=gray!20,inner sep=0pt,minimum size=15pt] (n2b) at (5,-1) {\small{$g_2$}};
\node[draw,circle,fill=gray!20,inner sep=0pt,minimum size=15pt] (n3b) at (5,-2) {\small{$g_3$}};
\node[draw,circle,fill=gray!20,inner sep=0pt,minimum size=15pt] (n4b) at (5,-3) {\small{$g_4$}};

\node[draw,circle,fill=gray!20,inner sep=0pt,minimum size=15pt] (n6b) at (5,-4) {\small{$g_5$}};
\node[draw,circle, fill=gray!80,inner sep=0pt,minimum size=15pt] (nRb) at (5,-5) {\small{R}};

\node[color=blue!60] () at (5.4,-1.75){\small{$e_1$}};
\node[color=blue!60] () at (5.5,-2.75){\small{$e_2$}};
\node[] () at (4.5,-3.75){\small{$e_0$}};

\path[-latex] (n1b) edge (n2b);
\path[-latex] (n2b) edge (n3b);
\path[-latex] (n3b) edge (n4b);
\path[-latex] (n4b) edge (n6b);
\path[-latex] (n6b) edge (nRb);

\path[-latex,out=240,in=135 ] (n1b) edge (n6b);
\path[-latex,out=330,in=60, dashed,blue!60 ] (n1b) edge (n3b);
\path[-latex,out=0,in=45,dashed,blue!60 ] (n1b) edge (n4b);

\end{scope}
\end{tikzpicture}
\label{fig:hans}
\caption{Left: Original DAG $\mathcal{D}_a$ with corresponding set of hierarchical relationship classes $A^{\mathcal{D}_a}$;
Right: Reconstruction $\hat{\mathcal{D}}_a$ of DAG $\mathcal{D}_a$ based on $A^{\mathcal{D}_a}$}

\end{figure} 
Given DAG $\mathcal{D}_a = \Big\{\mathcal{G}_{\mathcal{D}_a},\mathcal{E}_{\mathcal{D}_a}  \Big\}$ as displayed on the LHS of Fig.~5 and the corresponding set of hierarchical relationship classes represented by the corresponding set of selection variables $\text{A}^{\mathcal{D}_a}$. Note that $\alpha_4^{\mathcal{D}_a}(1,2) = \alpha_4^{\mathcal{D}_a}(1,3) = \alpha_4^{\mathcal{D}_a}(1,4) = 1$ due to edge $e_0 \in \mathcal{E}_{\mathcal{D}_a}$ and $\alpha_1^{\mathcal{D}_a}(1,5) = 1$. Now assume that we want to compute the topology of DAG $\mathcal{D}_a$, i.e., the set of edges $\mathcal{E}_{\mathcal{D}_a}$, based on $\text{A}^{\mathcal{D}_a}$.
DAG $\hat{\mathcal{D}}_a$ displayed on the RHS of Fig.~5 shows the estimated topology of DAG $\mathcal{D}_a$ based on $\text{A}^{\mathcal{D}_a}$. 
It can be shown that the black edges of DAG $\hat{\mathcal{D}}_a$ are necessary such that DAG $\hat{\mathcal{D}}_a$ is represented by $\text{A}^{\mathcal{D}_a}$. Edges $e_1$ and $e_2$ in DAG $\hat{\mathcal{D}}_a$ are optional in a sense that their existence has no effect on set $\text{A}^{\mathcal{D}_a}$. Edges $e_1$ and $e_2$ create two paths form gene $g_1$ to the reporter node $R$ which are independent of gene $g_2$ and gene $g_3$ respectively. However, due to edge $e_0$ gene $g_1$ already has a path to the reporter node $R$ which is independent of genes $g_2$ and $g_3$. Since $\alpha_4^{\mathcal{D}_a}(1,2) = \alpha_4^{\mathcal{D}_a}(1,3) = \alpha_4^{\mathcal{D}_a}(1,4) = 1$ do not contain information on the number of paths from gene $g_1$ to $R$ that are independent of $g_2$, $g_3$ and $g_4$, edges $e_1$ and $e_2$ do not affect the pattern of hierarchical relationship classes representing DAG $\mathcal{D}_a$, i.e., $\text{A}^{\mathcal{D}_a}$ 
and hence this yields ambiguities in computing the topology of DAG $\mathcal{D}_a$ based on its corresponding set of selection variables $\text{A}^{\mathcal{D}_a}$.         
%
Since it is a common assumption in genomics research, that GI-maps, i.e., DAGs, are not overly dense but rather sparse, we propose a policy which computes the sparsest DAG topology based on the detected pattern of hierarchical relationship classes. Given the detected pattern of hierarchical relationship classes of a DAG $\mathcal{D}$, i.e., $\hat{\text{A}}^{\mathcal{D}} =\bigcup\limits_{ \forall i,j \in \mathcal{G}: j>i }   \Big\{ \hat{\alpha}_1^{\mathcal{D}}(i,j),..., \hat{\alpha}_5^{\mathcal{D}}(i,j) \Big\} $, we compute an estimate $\hat{\mathcal{E}}_{\mathcal{D}}$ of the true topology set $\mathcal{E}_{\mathcal{D}}$ of DAG $\mathcal{D}$ according to the policy depicted in Tab.~\ref{table:edge_detection} 
where we make use of the symmetry properties $\hat{\alpha}_1^{\mathcal{D}}(i,j) = \hat{\alpha}_2^{\mathcal{D}}(j,i)$, $\hat{\alpha}_2^{\mathcal{D}}(i,j) = \hat{\alpha}_1^{\mathcal{D}}(j,i)$, $\hat{\alpha}_3^{\mathcal{D}}(i,j) = \hat{\alpha}_3^{\mathcal{D}}(j,i)$, $\hat{\alpha}_4^{\mathcal{D}}(i,j) = \hat{\alpha}_5^{\mathcal{D}}(j,i)$ and $\hat{\alpha}_5^{\mathcal{D}}(i,j) = \hat{\alpha}_4^{\mathcal{D}}(j,i)$ to redundantly expand the set of detected selection variables
in order to obtain a compact formulation of the mutually exclusive conditions $\text{E}_1$ to $\text{E}_4$ as stated in Tab.~\ref{table:edge_detection}. \\
Assume that either condition $\text{E}_1$ or condition $\text{E}_2$ is fulfilled, then we conclude that there is an edge from gene $i$ to gene $j$ in DAG $\mathcal{D}$. Given that either condition $\text{E}_3$ or condition $\text{E}_4$ is fulfilled, we conclude that there exists an edge from gene $j$ to gene $i$ in DAG $\mathcal{D}$. We remark that there cannot be an edge between two genes $i,j \in$$ \mathcal{G}:j>i$ if they are independent of each other, i.e., $\hat{\alpha}_3^{\mathcal{D}}(i,j)=1$. 
 \begin{table}[h!]

\begin{framed}
\begin{enumerate}
\item[$$] \begin{center} {\bf{Detection Policy:}} Compute sparsest DAG in line with $\hat{\text{A}}^{\mathcal{D}}$ \end{center} \vspace{1em}
\item[$\text{E}_1$:]{
\begin{align*}
\begin{split}
\hspace{-3em}	\bigg(  \hat{\alpha}_1^{\mathcal{D}}(i,j)=1 \bigg) \bigwedge \bigg(  \nexists l \in \mathcal{G} \setminus \left(i,j\right): \\
 \hat{\alpha}_1^{\mathcal{D}}(i,l)=1 \bigwedge \hat{\alpha}_2^{\mathcal{D}}(j,l)=1   \bigg)
	\end{split}
	\end{align*}
	
} 
\item[$\bullet$] $\Longrightarrow$ there is an edge in DAG $\mathcal{D}$ from gene $i$ to gene $j$, i.e $\left\{ i,j\right\}$

\item[$\text{E}_2$:]{
\begin{align*}
\begin{split}	\bigg(  \hat{\alpha}_4^{\mathcal{D}}(i,j)=1 \bigg) \bigwedge \bigg(  \nexists l \in \mathcal{G} \setminus \left(i,j\right): \\
\Big( \hat{\alpha}_1^{\mathcal{D}}(i,l)=1  \bigvee \hat{\alpha}_4^{\mathcal{D}}(i,l)=1 \Big) \bigwedge \\
 \Big( \hat{\alpha}_2^{\mathcal{D}}(j,l)=1 \bigvee \hat{\alpha}_5^{\mathcal{D}}(j,l)=1 \Big)  \bigg)
\end{split}
	\end{align*} 
} 
\item[$\bullet$] $\Longrightarrow$ there is an edge in DAG $\mathcal{D}$ from gene $i$ to gene $j$, i.e $\left\{ i,j\right\}$
\item[$\text{E}_3$:]{
\begin{align*}
\begin{split}
\bigg(  \hat{\alpha}_2^{\mathcal{D}}(i,j)=1  \bigg) \bigwedge \bigg(  \nexists l \in \mathcal{G} \setminus \left(i,j\right): \\
\hat{\alpha}_2^{\mathcal{D}}(i,l)=1 \bigwedge \hat{\alpha}_1^{\mathcal{D}}(j,l)=1   \bigg)	
		\end{split}
	\end{align*}
} 
\item[$\bullet$] $\Longrightarrow$ there is an edge in DAG $\mathcal{D}$ from gene $j$ to gene $i$, i.e $\left\{ j,i\right\}$

\item[$\text{E}_4$:]{
\begin{align*}
\begin{split}	\bigg(  \hat{\alpha}_5^{\mathcal{D}}(i,j)=1 \bigg) \bigwedge \bigg(  \nexists l \in \mathcal{G} \setminus \left(i,j\right): \\
\Big( \hat{\alpha}_2^{\mathcal{D}}(i,l)=1  \bigvee \hat{\alpha}_5^{\mathcal{D}}(i,l)=1 \Big) \bigwedge \\
 \Big( \hat{\alpha}_1^{\mathcal{D}}(j,l)=1 \bigvee \hat{\alpha}_4^{\mathcal{D}}(j,l)=1 \Big)  \bigg)
\end{split}
	\end{align*} 
} 
\item[$\bullet$] $\Longrightarrow$ there is an edge in DAG $\mathcal{D}$ from gene $j$ to gene $i$, i.e $\left\{ j,i\right\}$

\end{enumerate} 
\end{framed}

\caption{Proposed sparse edge detection policy}
\label{table:edge_detection}

\end{table}
\\
As described by $\text{E}_1$ there is an edge from gene $i$ to gene $j$ in DAG $\mathcal{D}$, if gene $i$ is linearly upstream of gene $j$, i.e., $\hat{\alpha}_1^{\mathcal{D}}(i,j)=1$, and there is no gene $l$ in DAG $\mathcal{D}$ that is linearly downstream of gene $i$, i.e., $\hat{\alpha}_1^{\mathcal{D}}(i,l)=1$, and linearly upstream of gene $j$, i.e., $\hat{\alpha}_2^{\mathcal{D}}(j,l)=1$.
According to condition $\text{E}_2$ there is an edge from gene $i$ to gene $j$ in DAG $\mathcal{D}$, if gene $i$ is upstream of gene $j$ with at least one path from gene $i$ to $R$ which is independent of gene $j$, and furthermore there is no gene $l$ in DAG $\mathcal{D}$ that is either linearly downstream of gene $i$ or downstream of gene $i$ with gene $i$ having at least one path to $R$ that is independent of $l$, and, gene $l$ is neither linearly upstream of gene $j$ nor is gene $l$ upstream of gene $j$ with an independent path to $R$. In order to elucidate the effect of condition $\text{E}_2$ onto the edge computation, we briefly turn to DAG $\hat{\mathcal{D}}_a$ in Fig.~5. Condition $\text{E}_2$ ensures that the optional edges $e_1$ and $e_2$ are not detected but only the necessary edges displayed in black color.      
We remark that conditions $\text{E}_3$ and $\text{E}_4$ can be elucidated by the same line of argument as used for conditions $\text{E}_1$ and $\text{E}_2$, but due to space limitations we omit a detailed explanation at this point.   
Finally, we propose a condition from which all reporter node edges, i.e, all egdes that connect gene $i \in \mathcal{G}$ with reporter node $R$ in DAG $\mathcal{D}$ can be computed.
Based on the detected set of hierarchical relationship classes, i.e., $\hat{\text{A}}^{\mathcal{D}}$, we follow our policy of computing the necessary edges only. For clarity of presentation and notational compactness, we define set $\mathcal{M}_i$ as
\begin{align}
\mathcal{M}_i = \left\{ l \in \mathcal{G} \setminus i | \ \hat{\alpha}_4^{\mathcal{D}}(i,l) = 1 \right\}, \quad i=1,...,G
\label{mi}
\end{align} 
containing all genes $l \in \mathcal{G}$ which are in class $k=4$ with gene $i \in \mathcal{G}$, i.e., $\hat{\alpha}_4^{\mathcal{D}}(i,l) = 1$.           
Furthermore, we define set $\mathcal{M}_i^{'}$ as 
\begin{align}
\mathcal{M}_i^{'} = \left\{ l \in \mathcal{M}_i | \ \exists \tilde{l} \in \mathcal{M}_i \setminus l:  \hat{\alpha}_3^{\mathcal{D}}(l,\tilde{l}) = 1 \right\}, \quad i=1,...,G
\label{mi_prime}
\end{align}
which contains all genes $l$ of set $\mathcal{M}_i$ that are independent of at least one other gene of set $\mathcal{M}_i$.
Based on sets $\mathcal{M}_i$ and $\mathcal{M}_i^{'}$, we formulate condition $\text{E}_R$ as stated in Tab.~\ref{table:edge_detection_reporter}.
\begin{table}[h!]

\begin{framed}
\begin{enumerate}
\item[$\text{E}_R$:]{
\begin{align*}
\begin{split}
\underbrace{\bigg(  \nexists l: \hat{\alpha}_1^{\mathcal{D}}(i,l) = 1 \bigg)}_{= \text{LHS}} \bigwedge \underbrace{\bigg( \mathcal{M}_i = \emptyset \bigvee \mathcal{M}_i^{'} = \emptyset  \bigg) }_{= \text{RHS}} 
 	\end{split}
	\end{align*}	
} 
 \end{enumerate} 
\end{framed}

\caption{Proposed reporter node edge detection policy}
\label{table:edge_detection_reporter}

\end{table}
We conclude that there is an edge from gene $i$ to reporter node $R$ in DAG $\mathcal{D}$, if condition $\text{E}_R$ is fulfilled. 
Given that gene $i$ is linearly upstream of at least a single gene $l$, i.e., $\hat{\alpha}_1^{\mathcal{D}}(i,l) = 1$, there cannot exist an edge from gene $i$ to reporter node $R$ in DAG $\mathcal{D}$, since all paths from gene $i$ to $R$ pass through at least one other gene $l$. Conversely, if there is no such gene $l$ that $\hat{\alpha}_1^{\mathcal{D}}(i,l) = 1$, then the LHS of $\text{E}_R$ as given in Tab.~\ref{table:edge_detection_reporter} is fulfilled.       
The RHS of $\text{E}_R$ accounts for our policy of detecting sparse DAGs only and is fulfilled if either $\mathcal{M}_i$, $\mathcal{M}_i^{'}$ or $\mathcal{M}_i$ and $\mathcal{M}_i^{'}$ are empty. Note that given $\mathcal{M}_i = \emptyset$ it follows that $\mathcal{M}_i^{'} = \emptyset$ as well, whereas the opposite is not true.  
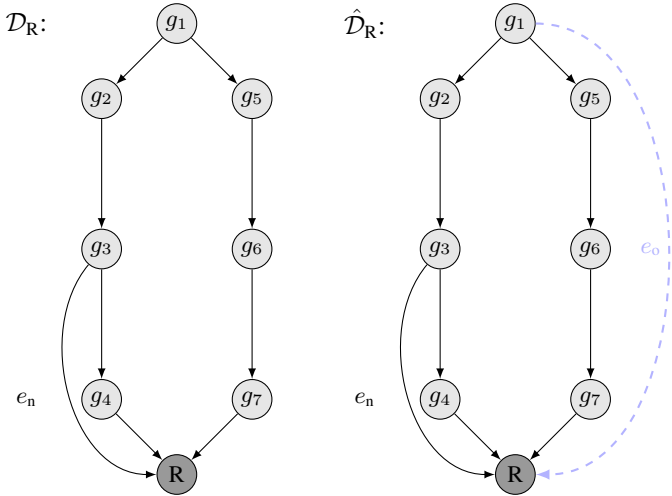
\begin{figure}[h!]
\centering
\begin{tikzpicture}

\begin{scope}[xshift=-1cm,yshift=0cm, scale=1]


\node[] () at (-1,0) {$\mathcal{D}_{\text{R}}$:};
\node[draw,circle,fill=gray!20,inner sep=0pt,minimum size=15pt] (n1) at (1,0) {\small{$g_1$}};

\node[draw,circle,fill=gray!20,inner sep=0pt,minimum size=15pt] (n2) at (0,-1) {\small{$g_2$}};
\node[draw,circle,fill=gray!20,inner sep=0pt,minimum size=15pt] (n3) at (0,-3) {\small{$g_3$}};
\node[draw,circle,fill=gray!20,inner sep=0pt,minimum size=15pt] (n4) at (0,-5) {\small{$g_4$}};

\node[draw,circle,fill=gray!20,inner sep=0pt,minimum size=15pt] (n2b) at (2,-1) {\small{$g_5$}};
\node[draw,circle,fill=gray!20,inner sep=0pt,minimum size=15pt] (n3b) at (2,-3) {\small{$g_6$}};
\node[draw,circle,fill=gray!20,inner sep=0pt,minimum size=15pt] (n4b) at (2,-5) {\small{$g_7$}};

\node[draw,circle, fill=gray!80,inner sep=0pt,minimum size=15pt] (nR) at (1,-6) {\small{R}};

\path[-latex] (n1) edge (n2);
\path[-latex] (n1) edge (n2b);

\path[-latex] (n2) edge (n3);
\path[-latex] (n3) edge (n4);
\path[-latex] (n4) edge (nR);

\path[-latex] (n2b) edge (n3b);
\path[-latex] (n3b) edge (n4b);
\path[-latex] (n4b) edge (nR);

\path[-latex,out=230,in=180] (n3) edge (nR);
\node () at (-1,-5){ \small{$e_{\text{n}}$}};
\end{scope}
\begin{scope}[xshift=3.5cm,yshift=0cm, scale=1]
\node[] () at (-1,0) {$\hat{\mathcal{D}}_{\text{R}}$:};
\node[draw,circle,fill=gray!20,inner sep=0pt,minimum size=15pt] (n1) at (1,0) {\small{$g_1$}};

\node[draw,circle,fill=gray!20,inner sep=0pt,minimum size=15pt] (n2) at (0,-1) {\small{$g_2$}};
\node[draw,circle,fill=gray!20,inner sep=0pt,minimum size=15pt] (n3) at (0,-3) {\small{$g_3$}};
\node[draw,circle,fill=gray!20,inner sep=0pt,minimum size=15pt] (n4) at (0,-5) {\small{$g_4$}};

\node[draw,circle,fill=gray!20,inner sep=0pt,minimum size=15pt] (n2b) at (2,-1) {\small{$g_5$}};
\node[draw,circle,fill=gray!20,inner sep=0pt,minimum size=15pt] (n3b) at (2,-3) {\small{$g_6$}};
\node[draw,circle,fill=gray!20,inner sep=0pt,minimum size=15pt] (n4b) at (2,-5) {\small{$g_7$}};

\node[draw,circle, fill=gray!80,inner sep=0pt,minimum size=15pt] (nR) at (1,-6) {\small{R}};

\path[-latex] (n1) edge (n2);
\path[-latex] (n1) edge (n2b);

\path[-latex] (n2) edge (n3);
\path[-latex] (n3) edge (n4);
\path[-latex] (n4) edge (nR);

\path[-latex] (n2b) edge (n3b);
\path[-latex] (n3b) edge (n4b);
\path[-latex] (n4b) edge (nR);

\path[-latex,out=0,in=0, dashed,blue!30, thick ] (n1) edge (nR);
\node[color=blue!30] () at (2.8,-3){ \small{$e_{\text{o}}$}};

\path[-latex,out=230,in=180] (n3) edge (nR);
\node () at (-1,-5){ \small{$e_{\text{n}}$}};
\end{scope}

\end{tikzpicture}
\label{fig:reporterNode_example}
\caption{Example DAG $\mathcal{D}_{\text{R}}$ to elucidate the functionality of the RHS of condition $\text{E}_{R}$ of Tab.~\ref{table:edge_detection_reporter} }

\end{figure} 
In order to explain the effect of the RHS of condition $\text{E}_R$ in an intuitive manner, let us turn to DAG $\mathcal{D}_{\text{R}}$ as displayed in Fig.~6.
Assume we are given the pattern of hierarchical relationship classes that corresponds to DAG $\mathcal{D}_{\text{R}}$, i.e., $A^{\mathcal{D}_{\text{R}}}$ and we want to compute all reporter node edges based on $A^{\mathcal{D}_{\text{R}}}$, i.e., all edges that directly connect a gene in DAG $\mathcal{D}_{\text{R}}$ with the reporter node $R$. 
Note that $\alpha_1^{\mathcal{D}_{\text{R}}}(2,3) = 1$, $\alpha_5^{\mathcal{D}_{\text{R}}}(2,4) = 1$, $\alpha_4^{\mathcal{D}_{\text{R}}}(3,4) = 1$ and $\alpha_4^{\mathcal{D}_{\text{R}}}(1,l) = 1 \ \forall l\in \mathcal{G}_{\mathcal{D}_R} \setminus \left\{ g_1, R\right\}$.
It can be shown that gene $g_3$ fulfills the LHS of $\text{E}_R$, i.e., there is no gene which is linearly downstream of $g_3$ and furthermore $\mathcal{M}_3 = \left\{ g_4 \right\}$ and $\mathcal{M}_3^{'} = \emptyset$. Hence, condition $\text{E}_R$ is fulfilled and edge $e_n$ connecting $g_3$ and $R$ is computed in DAG $\hat{\mathcal{D}}_{\text{R}}$ that is the reconstruction of DAG $\mathcal{D}_{\text{R}}$ based on $A^{\mathcal{D}_{\text{R}}}$. Furthermore, for set $A^{\mathcal{D}_{\text{R}}}$ the edge $e_n$ in the reconstructed DAG $\hat{\mathcal{D}}_{\text{R}}$ is necessary, since $\alpha_1^{\mathcal{D}_{\text{R}}}(2,3) = 1$, $\alpha_5^{\mathcal{D}_{\text{R}}}(2,4) = 1$ and $\alpha_4^{\mathcal{D}_{\text{R}}}(3,4) = 1$. In contrast to $e_n$ edge $e_o$ is not necessary for $A^{\mathcal{D}_{\text{R}}}$ to represent $\hat{\mathcal{D}}_{\text{R}}$, since $\alpha_4^{\mathcal{D}_{\text{R}}}(1,l) = 1 \ \forall l\in \mathcal{G}_{\mathcal{D}_R} \setminus \left\{ g_1, R\right\}$ irrespectively of edge $e_o$. Hence, $e_o$ is not detected, since $\mathcal{M}_1 \neq \emptyset$ and $\mathcal{M}_1^{'} \neq \emptyset$.                  
\\
\\         
We obtain an estimate $\mathcal{E}_{\text{GENIE}}$ of the true set of edges $\mathcal{E}_{\mathcal{D}}$ of DAG $\mathcal{D}$ by setting $\hat{\text{A}}^{\mathcal{D}} = A^{\text{O}_{\text{GENIE}}}$ and evaluating conditions $\text{E}_1$ to $\text{E}_4$ and condition $\text{E}_R$ as stated in Tab.~\ref{table:edge_detection} and Tab.~\ref{table:edge_detection_reporter}, respectively.

\section{GI-GENIE-Algorithm}
In this section, we extend program $\text{O}_{\text{GENIE}}$ in (\ref{parent_genie}) by incorporating GI-profile data, in order to jointly detect the most consistent pattern of hierarchical relationship classes and the corresponding set of edges $\mathcal{E}_{\text{GI}}$ that is an estimate of the true graph topology of DAG $\mathcal{D}$, i.e., the set of edges $\mathcal{E}_{\mathcal{D}}$ which underlies the SK, DK and GI-profile data. 
\\
Let us define the following selection variables
\begin{align}
\beta(i,j) = \begin{cases} 1 & \exists \ \text{edge between} \ $i,j$\\ 0 & \text{no edge} \end{cases} \quad \forall i,j \in \mathcal{G}:j>i 
\label{beta}
\end{align}
where $\beta(i,j) =1$ denotes that there is an edge between genes $i,j \in \mathcal{G}:j>i $ in DAG $\mathcal{D}$ and $\beta(i,j) =0$ denotes that there exists no edge between genes $i$ and $j$. Note that unlike $\alpha_k(i,j)=1$ for $k \in \mathcal{K}$, $\beta(i,j) =1$ does not capture directionality information 
about the graph topology, i.e., $\beta(i,j) = 1$ states that there is an edge between genes $i,j$ in DAG $\mathcal{D}$ without specifying the hierarchy among both genes.
The GI-GENIE program of jointly detecting the most consistent pattern of hierarchical relationship classes $\text{A}^{\text{O}_{\text{GI-GENIE} }}$ and the corresponding DAG topology $\mathcal{E}_{\text{GI}}$ based on SK, DK and GI-profile data can be formulated as linear integer program  
		\begin{subequations}
		\begin{align} 
		\text{O}_{\text{GI-GENIE}}: & \nonumber \\
		  \min_{\left\{ \alpha_k(i,j), \beta(i,j), z_l(i,j) \right\}} \; & \quad  \lambda_d \sum\limits_{i=1}^{G} \sum\limits_{j=i+1}^{G}  \Big( \sum\limits_{k=1}^{\left| \mathcal{K} \right|}    s_k(i,j)\alpha_k(i,j) \Big) \nonumber \\ 
		& -\lambda_s  \sum\limits_{i=1}^{G} \sum\limits_{j=i+1}^{G}  \Big( \sum\limits_{l}    z_l(i,j) \Big) \nonumber \\
	&	-	\lambda_c \sum\limits_{i=1}^{G} \sum\limits_{j=i+1}^{G}   \rho(i,j)\beta(i,j) \nonumber \\ 
	&+ \lambda_p \sum\limits_{i=1}^{G} \sum\limits_{j=i+1}^{G}  \beta(i,j) \label{objective_gi_genie} \\	
 \text{s. t.:} \hspace{2em} \; &  \text{Eq}.~(\text{\ref{alpha_genie}})  \text{- Eq}.~(\text{\ref{consistency_set}}) \\
			& \beta(i,j) \in \left\{0,1 \right\} \ \forall i,j \in \mathcal{G}: j>i \\
			& z_l(i,j) \in \left\{0,1 \right\} \  \forall l \in \mathcal{G} \setminus \left\{i,j\right\}, \nonumber \\
			& \hspace{2em} \forall i,j \in \mathcal{G}: j>i \\
			& 1- \alpha_3(i,j) \geq \beta(i,j)  \nonumber \\
			& \forall i,j \in \mathcal{G}: j>i \label{C1} \\
			& \mathcal{L}_c \Longrightarrow \ \text{additional topology constraints} \label{C2} \\
			& \left| \mathcal{G} \right|-2+ \beta(i,j) \geq 1+ \sum\limits_{l \in \mathcal{G} \setminus \left\{ i,j\right\}} z_l(i,j) \label{C3}  \\
			& \forall i,j \in \mathcal{G}: j>i  \nonumber
		\end{align}
		\end{subequations}
where $\rho(i,j)$ denotes a measure of the similarity between the GI-profiles of genes $i,j \in \mathcal{G}:j>i$ according to (\ref{GI_data}), the scalars $\lambda_d,\lambda_s,\lambda_c,\lambda_p$ are non-negative weighting constants and $z_l(i,j) \forall i,j,l \in \mathcal{G}: j>i, l\neq i, l\neq j$ are binary slack variables. Note that $\lambda_s$ is assumed to be a very small non-negative constant, i.e. $0 \leq \lambda_s \ll 1 $.
The GI-profile (GIP) term 
\begin{align}
-	\lambda_c \sum\limits_{i=1}^{G} \sum\limits_{j=i+1}^{G}   \rho(i,j)\beta(i,j) + \lambda_p \sum\limits_{i=1}^{G} \sum\limits_{j=i+1}^{G}  \beta(i,j) 
\label{GIP}
\end{align}
in the objective function (\ref{objective_gi_genie}) of program $\text{O}_{\text{GI-GENIE}}$ seeks to detect an edge between genes $i,j \in \mathcal{G}:j>i$ if the GI-profile similarity $\rho(i,j)$ is greater than the predefined threshold $\frac{\lambda_p}{\lambda_c}$. The slack term 
\begin{align}
-\lambda_s  \sum\limits_{i=1}^{G} \sum\limits_{j=i+1}^{G}  \Big( \sum\limits_{l}    z_l(i,j) \Big)
\label{slack}
\end{align}
is necessary for the correct functionality of constraints (\ref{C2}) and (\ref{C3}) as explained below.
The slack variables $z_l(i,j) \forall i,j,l \in \mathcal{G}: j>i, l\neq i, l\neq j$ are generally necessary to ensure that the information about the topology of DAG $\mathcal{D}$, which is encoded in the pattern of selection variables $\text{A}^{\text{O}_{\text{GI-GENIE} }}$ detected by program $\text{O}_{\text{GI-GENIE}}$, is not contradicting with the set of edge selection variables $ \left\{ \hat{\beta}(i,j) \right\}  \, \forall i,j \in \mathcal{G}:j>i$ detected by program $\text{O}_{\text{GI-GENIE}}$.
In particular, given that the detected pattern of selection variables $\text{A}^{\text{O}_{\text{GI-GENIE} }}$ enforces that there is an edge between genes $i,j$ in DAG $\mathcal{D}$, then the slack variables ensure that the corresponding edge selection variable indicates that there is an edge between genes $i,j$, i.e., $\hat{\beta}(i,j) = 1$.
Furthermore, given that the detected pattern of selection variables $\text{A}^{\text{O}_{\text{GI-GENIE} }}$ enforces that there is no edge between genes $i,j$ in DAG $\mathcal{D}$, then the slack variables ensure that the corresponding edge selection variable indicates that there is no edge between genes $i,j$, i.e., $\hat{\beta}(i,j) = 0$.  
On the contrary, assume that the detected edge selection variables enforce that there is an edge between genes $i,j$ in DAG $\mathcal{D}$, i.e., $\hat{\beta}(i,j) = 1$, then 
the $z_l(i,j) \forall i,j,l \in \mathcal{G}: j>i, l\neq i, l\neq j$ ensure that the detected pattern of selection variables $\text{A}^{\text{O}_{\text{GI-GENIE} }}$ must fulfill one of the conditions stated in Tab.~\ref{table:edge_detection}.
Consequently, given that the detected edge selection variables enforce that there is no edge between genes $i,j$ in DAG $\mathcal{D}$, i.e., $\hat{\beta}(i,j) = 0$, then 
the $z_l(i,j) \forall i,j,l \in \mathcal{G}: j>i, l\neq i, l\neq j$ ensure that the detected pattern of selection variables $\text{A}^{\text{O}_{\text{GI-GENIE} }}$ does not 
fulfill any of the conditions stated in Tab.~\ref{table:edge_detection}.
\\
Furthermore, let the auxiliary variables
\begin{align}
q(i,j) = \begin{cases} 1 & \rho(i,j) \geq  \frac{\lambda_c}{\lambda_p} \\ 0 & \rho(i,j) < \frac{\lambda_c}{\lambda_p}   \end{cases} \quad \forall \, i,j \in \mathcal{G}:j>i
\end{align}		
describe the detection of the edges of DAG $\mathcal{D}$ based on GI-profile data only, where $q(i,j) = 1$ denotes that there is an edge between genes $i,j \in \mathcal{G}:j>i$ and $q(i,j)=0$ denotes that there is no edge between genes $i,j \in \mathcal{G}:j>i$.
Since any pattern of hierarchical relationship classes implies a specific set of edges and any set of edges implies a specific pattern of hierarchical relationship classes, there is a strong coupling of constraints, i.e. there are strong logical implications
among the selection variables $\alpha_k(i,j) \ \forall \, i,j \in \mathcal{G}:j>i, \ \forall \, k \in \mathcal{K}$ and the selection variables $\beta(i,j) \ \forall \, i,j \in \mathcal{G}:j>i$, the constraints in Eq.~(\ref{C1}) to Eq.~(\ref{C3}) ensure that the detected hierarchical relationship classes and the corresponding edges, i.e., the $\alpha_k(i,j)$ and $\beta(i,j)$, are not mutually contradicting.
Given that two genes $i,j \in \mathcal{G}:j>i$ in DAG $\mathcal{D}$ are independent, i.e., $\alpha_3(i,j) = 1$, there cannot exist an edge between those genes in DAG $\mathcal{D}$, i.e., $\beta(i,j)=0$. This logical implication is reflected by (\ref{C1}). 
Set $\mathcal{L}_c$ in (\ref{C2}) and the linear integer inequalities in (\ref{C3}) model conditions $\text{E}_1$ to $\text{E}_4$ of our proposed edge detection policy as stated in Tab.~\ref{table:edge_detection}.
Since we do not want to redundantly expand the set of selection variables $\alpha_k(i,j)$ to all $i,j \in \mathcal{G}: j \neq i$ in order to not increase the complexity of program $\text{O}_{\text{GI-GENIE}}$, we have to consider three cases when formulating conditions $\text{E}_1$ to $\text{E}_4$ of Tab.~\ref{table:edge_detection} as linear integer inequalities, i.e., $i,j,l \in \mathcal{G}: l>j>i$, $i,j,l \in \mathcal{G}: j>i>l$ and $i,j,l \in \mathcal{G}: j>l>i$. 
Then the constraints in set $\mathcal{L}_{c,1}$
\begin{subequations}
\begin{align}
\mathcal{L}_{c,1} = \bigg\{ & \nonumber \\
& 1 -\beta(i,j)  \geq \alpha_1(i,j) + \alpha_1(i,l) + \alpha_2(j,l) -2  \label{e1} \\
& 1 -z_l(i,j)  \geq \alpha_1(i,j) + \alpha_1(i,l) + \alpha_2(j,l) -2  \label{z1} \\
& \frac{1}{2} \left( \alpha_1(i,l) + \alpha_2(j,l)    \right) \geq \alpha_1(i,j) - z_l(i,j)   \label{z1_mod} \\
& \nonumber \\
& 1 -\beta(i,j)  \geq \alpha_2(i,j) + \alpha_2(i,l) + \alpha_1(j,l) -2 \nonumber   \\
& 1 -z_l(i,j)  \geq \alpha_2(i,j) + \alpha_2(i,l) + \alpha_1(j,l) -2 \nonumber \\
& \frac{1}{2} \left( \alpha_2(i,l) + \alpha_1(j,l)    \right) \geq \alpha_2(i,j) - z_l(i,j) \label{e2} \\
& \nonumber \\
& 1 -\beta(i,j) + q(i,j) \geq \alpha_4(i,j) + \alpha_1(i,l) +  \nonumber  \\
& \alpha_4(i,l) + \alpha_2(j,l) +\alpha_5(j,l)  -2  \label{e4_q0} \\
& 1 -z_l(i,j) + q(i,j) \geq \alpha_4(i,j) + \alpha_1(i,l) +  \nonumber \\
& \alpha_4(i,l) + \alpha_2(j,l) +\alpha_5(j,l)  -2 \label{z4_q0} \\
& \frac{1}{2} \left( \alpha_1(i,l) + \alpha_4(i,l) + \alpha_2(j,l)+\alpha_5(j,l)\right) \geq \nonumber \\
& \alpha_4(i,j) - z_l(i,j) - q(i,j) \label{z4_q0_mod} \\
& \nonumber \\
& 2 -\beta(i,j) - q(i,j) \geq \alpha_4(i,j) + \alpha_1(i,l) \nonumber  \\
& + \alpha_2(j,l)  + \alpha_5(j,l) -2   \label{e4_q1} \\
& 2 -z_l(i,j) - q(i,j) \geq \alpha_4(i,j) + \alpha_1(i,l) \nonumber  \\
& + \alpha_2(j,l) +\alpha_5(j,l) -2   \label{z4_q1}\\
& \frac{1}{2} \left( \alpha_1(i,l) + \alpha_2(j,l)\right) \geq \nonumber \\
& \alpha_4(i,j) - z_l(i,j) -1+ q(i,j) \label{z4_q1_mod} \\
& \nonumber \\
& 1 -\beta(i,j) + q(i,j) \geq \alpha_5(i,j) + \alpha_2(i,l) +  \nonumber  \\
& \alpha_5(i,l) + \alpha_1(j,l) +\alpha_4(j,l)  -2 \nonumber \\
& 1 -z_l(i,j) + q(i,j) \geq \alpha_5(i,j) + \alpha_2(i,l) +  \nonumber  \\
& \alpha_5(i,l) + \alpha_1(j,l) +\alpha_4(j,l)  -2 \nonumber \\
& \frac{1}{2} \left( \alpha_2(i,l) + \alpha_5(i,l) + \alpha_1(j,l)+\alpha_4(j,l)    \right) \geq \nonumber\\
& \alpha_5(i,j) - z_l(i,j) - q(i,j) \label{e5_q0} \\
& \nonumber \\
& 2 -\beta(i,j) - q(i,j) \geq \alpha_5(i,j) + \alpha_2(i,l) \nonumber \\
& + \alpha_5(i,l) + \alpha_1(j,l)  -2 \nonumber \\
& 2 -z_l(i,j) - q(i,j) \geq \alpha_5(i,j) + \alpha_2(i,l) \nonumber \\
& + \alpha_5(i,l) +\alpha_1(j,l)  -2 \nonumber \\
& \frac{1}{2} \left( \alpha_2(i,l) + \alpha_1(j,l) \right) \geq \nonumber \\
& \alpha_5(i,j) - z_l(i,j) -1+ q(i,j) \label{e5_q1} \\
& \bigg\} \forall i,j,l \in \mathcal{G}: l>j>i \nonumber 
\end{align} 
\end{subequations}
model the logical implications among the selection variables $\alpha_k(i,j), \alpha_{k'}(i,l), \alpha_{k''}(j,l) $ and $\beta(i,j)$ for $\  k,k',k'' \in \mathcal{K}$, $\forall i,j,l \in \mathcal{G}: l>j>i$. Together with (\ref{C3}) and the slack term in (\ref{slack}), constraints (\ref{e1})-(\ref{z1_mod}) model condition $\text{E}_1$ of our detection policy taking into account the GI-profile similarity information $\rho(i,j)$ via selection variables $\beta(i,j)$.
Assume that based on the SK and DK phenotypes it is most consistent that $\alpha_1(i,j)= \alpha_1(i,l)= \alpha_2(j,l) = 1$ for at least one gene $l$ in DAG $\mathcal{D}$ which corresponds to condition $\text{E}_1$ being violated. Hence, there cannot exist an edge between genes $i$ and $j$ in DAG $\mathcal{D}$. In this case the RHS of (\ref{e1}) amounts to $1$ which enforces the LHS of (\ref{e1}) to amount to $1$ as well, i.e., $\beta(i,j)=0$. Note that for $\alpha_1(i,j)= \alpha_1(i,l)= \alpha_2(j,l) = 1$, (\ref{z1_mod}) makes no restrictions on $z_l(i,j)$ but by (\ref{z1}) $z_l(i,j)$ is forced to $0$, thus (\ref{C3}) is fulfilled as well. 
Furthermore, assume for genes $i,j \in \mathcal{G}:j>i$, based on the SK and DK phenotypes, it is most consistent that $\alpha_1(i,j)=1$, but $\alpha_1(i,l)$ and $\alpha_2(j,l)$ are not jointly $1$ for all other genes $l \in \mathcal{G}: l>j>i $, i.e., $\alpha_1(i,l) + \alpha_1(j,l) < 2$, then there is an edge between genes $i,j$ in DAG $\mathcal{D}$ according to condition $\text{E}_1$. In this case it is obvious that (\ref{e1}) and (\ref{z1}) are always fulfilled, i.e., there are no restrictions on $\beta(i,j)$ and $z_l(i,j)$ by (\ref{e1}) and (\ref{z1}). However, due to the slack term in (\ref{slack}), it follows that $z_l(i,j) =1 \, \forall l \in \mathcal{G}:l > j >i$ and $\beta(i,j)$ is set to $1$ in order to fulfill (\ref{C3}).
\\
Given that the GI-profile similarity data strongly supports that there is no edge between genes $i,j$ in DAG $\mathcal{D}$, i.e., $\beta(i,j) = 0$, and $\alpha_1(i,j)=1$ is most consistent based on the SK and DK phenotypes measured, then it follows from (\ref{C3}) that there must be at least one $l \in \mathcal{G}: l > j>i$ for which $z_l(i,j)= 0$. In this case, with $\beta(i,j) = 0$, $\alpha_1(i,j)=1$ and $z_l(i,j)= 0$, the RHS of (\ref{z1_mod}) amounts to $1$, forcing the LHS of (\ref{z1_mod}) to amount to $1$ as well, i.e., $\alpha_1(i,l) = 1$ and $\alpha_2(j,l) = 1$, which is together with the assumption of $\alpha_1(i,j)=1$ a combination that violates the existence of a direct edge between genes $i$ and $j$. Furthermore, note that (\ref{e1}) and (\ref{z1}) do not have any implications on the selection variables $\alpha_1(i,j), \alpha_1(i,l), \alpha_2(j,l)$ for the case that $\beta(i,j)=0$ and $z_l(i,j)= 0$. \\
Assume that the GI-profile similarity data strongly supports that there is an edge between genes $i,j$ in DAG $\mathcal{D}$, i.e., $\beta(i,j) = 1$, and $\alpha_1(i,j)=1$ is most consistent based on the SK and DK phenotypes measured, then according to (\ref{e1}) there cannot be any gene $l \in \mathcal{G}: l >j>i$ for which $\alpha_1(i,l) = 1$ and $\alpha_2(j,l) = 1$. Hence, the RHS of (\ref{z1}) is smaller or equal to $0$ which enforces $z_l(i,j)=1 \, \forall l \in \mathcal{G}: l >j>i $ due to the slack term in (\ref{slack}). Thus, (\ref{C3}) is fulfilled with equality. Note that in this case (\ref{z1_mod}) is always fulfilled.       
Together with (\ref{C3}) and the slack term in (\ref{slack}), the three inequalities in (\ref{e2}) model condition $\text{E}_2$ where we can elucidate their functionality in the same fashion as before.
Constraints (\ref{e4_q0}) to (\ref{z4_q1_mod}) along with (\ref{C3}) and the slack term in (\ref{slack}) model a minor modification of condition $\text{E}_3$ where we not only detect all necessary edges, but also optional edges given that their existence is strongly supported by the GI-profile.
Given that the existence of an edge between genes $i,j \in \mathcal{G}:j>i$ in DAG $\mathcal{D}$ is not strongly supported by the GI-profile, i.e. $q(i,j)=0$, constraints (\ref{e4_q0}) to (\ref{z4_q0_mod}) along with (\ref{C3}) and (\ref{slack}) model condition $\text{E}_3$ which only allows necessary edges to be detected and we can elucidate their functionality in the same fashion as in (\ref{e1}) to (\ref{z1_mod}). Note that (\ref{e4_q1}) to (\ref{z4_q1_mod}) are always fulfilled for $q(i,j)=0$, i.e., no implications among the selection variables $\alpha_k(i,j)$ and $\beta(i,j)$ are posed.
Assuming that the existence of an edge between genes $i,j \in \mathcal{G}:j>i$ in DAG $\mathcal{D}$ is strongly supported by the GI-profile, i.e. $q(i,j)=1$, then the constraints in (\ref{e4_q0}) and (\ref{z4_q0_mod}) are always fulfilled, i.e., no implications among the selection variables $\alpha_k(i,j)$ and $\beta(i,j)$ are posed by (\ref{e4_q0}) and (\ref{z4_q0_mod}). However, constraints (\ref{e4_q1}) and (\ref{z4_q1_mod}) pose relaxed logical implications among the selection variables $\alpha_k(i,j)$ and $\beta(i,j)$ compared to constraints (\ref{e4_q0}) to (\ref{z4_q0_mod}). Hence given that $q(i,j)=1$ and $\alpha_4(i,j)=1$, an edge between genes $i,j \in \mathcal{G}:j>i$ in DAG $\mathcal{D}$ is detected if it is allowed by the pattern of hierarchical relationship classes.                           
Constraints (\ref{e5_q0}) to (\ref{e5_q1}) along with (\ref{C3}) and the slack term in (\ref{slack}) model a minor modification of condition $\text{E}_4$ where we not only detect all necessary edges, but also optional edges given that their existence is strongly supported by the GI-profile. Furthermore, the functionality of constraints (\ref{e5_q0}) to (\ref{e5_q1}) can be explained with the same line of argument as used to elucidate constraints (\ref{e4_q0}) to (\ref{z4_q1_mod}).
\\
Denote $\mathcal{L}_{c,2}$ and $\mathcal{L}_{c,3}$ as the sets of topology constraints that model the logical coupling among the selection variables $\alpha_k(i,j), \alpha_{k'}(i,l),\alpha_{k''}(j,l) $ and $\beta(i,j)$ for $k, k', k'' \in \mathcal{K}$ and $i,j,l \in \mathcal{G}: j>i>l$ and $i,j,l \in \mathcal{G}: j>l>i$ respectively.
Then the full set of coupled constraints of the selection variables $\alpha_k(i,j)$ and $\beta(i,j)$ is given by
\begin{align}
\mathcal{L}_c = \bigcup\limits_{\kappa=1}^{3} \left\{ \mathcal{L}_{c,\kappa} \right\} \label{L_c}  
\end{align}  
where we again refer the interested reader to \cite{supp_mat} for a detailed description of $\mathcal{L}_c$. 
We obtain an estimate $\mathcal{E}_{\text{GI}}$ of the true set of edges $\mathcal{E}_{\mathcal{D}}$ of DAG $\mathcal{D}$ based on the computed set of edge selection variables $\hat{\beta}(i,j)$ of program $\text{O}_{\text{GI-GENIE}}$ where we infer the directionality of the edges according to $\text{A}^{\text{O}_{\text{GI-GENIE}}}$ that is the most consistent pattern of hierarchical relationship classes given the observed SK, DK and GI-profile data. Note that all reporter node edges are computed according to our proposed reporter node edge detection policy as given in Tab.~\ref{table:edge_detection_reporter}. Since the reporter node is an artificial node in the concept of a DAG, there is no GI-profile similarity data $\rho(i,R) \, \forall i \in \mathcal{G}$ and thus, no edge selection variable $\beta(i,R) \, \forall i \in \mathcal{G}$ according to \ref{beta}.

\section{Sequential Scalability Technique}
Due to the combinatorial nature of problems $\text{O}^{\text{GENIE}}$ and $\text{O}^{\text{GI-GENIE}}$, the GENIE algorithm and GI-GENIE algorithm, respectively, cannot be applied to the data of large sets of genes $\mathcal{G}$, since the complexity of problems $\text{O}^{\text{GENIE}}$ and $\text{O}^{\text{GI-GENIE}}$ grows exponentially with the number of genes and both problems become NP-hard. 
In order to nevertheless obtain statistically stressable statements about the interactions among genes in a large set of genes $\mathcal{G}$, we propose the sequential-scalability (SEQSCA)-technique which is based on the GENIE-algorithm and the GI-GENIE algorithm, respectively.
\\
In particular, we create a sequence of $S$ subsets $\left\{ \mathcal{G}_s \right\}_{1}^{S}$ of the full set of genes $\mathcal{G}$, i.e., $\mathcal{G}_s \subset \mathcal{G}, \ \forall s \in \left\{1,...,S \right\}$, where we estimate the topology $\mathcal{E}_{\mathcal{D},s}$ of each DAG $\mathcal{D}_s$, underlying the data of the subset of genes $\mathcal{G}_s$, by the GENIE or GI-GENIE-algorithm, respectively, in order to translate the estimated topology $\mathcal{E}_{\mathcal{D},s}$ of DAG $\mathcal{D}_s$ into the corresponding adjacency matrix $\boldsymbol{M}_s$ for each $s \in \left\{1,..., S \right\}$. 
Based on the computed sequence of adjacency matrices $\left\{ \boldsymbol{M}_s \right\}_{1}^{S}$, we iteratively compute the reliability matrix $\boldsymbol{M} \in \left[ 0,1 \right]^{ N \times N}$ of the full set of genes $\mathcal{G}$ in such a way, that each entry $\left[ \boldsymbol{M} \right]_{i,j \in \mathcal{G}}$ describes the empirical probability of an edge to exist between genes $i,j \in \mathcal{G}$, i.e., the empirical probability that genes $i,j \in \mathcal{G}$ directly interact with each other, where a value of $0$ means that there is an interaction between the considered pair of genes with probability $0$ and a value of $1$ means that the considered pair of genes interacts with probability $1$. 
\\
In particular, in each iteration $s$ we consider a subset $\mathcal{G}_s$ of size $N_S \ll \left| \mathcal{G} \right|$ of the full set of genes $\mathcal{G}$, where each gene of $\mathcal{G}_s$ is selected from $\mathcal{G}$ without replacement with equal probability. 
Based on the selected subset $\mathcal{G}_s$, we compute in each iteration $s$ an estimate $\mathcal{E}_{\mathcal{D},s}$ of the true topology of DAG $\mathcal{D}_{s}$, underlying the observed data of the genes in subset $\mathcal{G}_s$, by the GENIE or the GI-GENIE-algorithm, respectively. Furthermore, the topology estimate $\mathcal{E}_{\mathcal{D},s}$ of DAG
$\mathcal{D}_s$ is translated into the corresponding adjacency matrix $\boldsymbol{M}_s$.
The update of the reliability matrix for iteration $s$ is computed according to Eq.~(\ref{superAD})   
\begin{align}
\left[ \boldsymbol{M}^{(s+1)}  \right]_{i,j} = \left[ \boldsymbol{M}^{(s)} \right]_{i,j} + \left[ \boldsymbol{M}_s \right]_{\kappa_i, \kappa_j} \quad \forall i,j \in \mathcal{G}_s
\label{superAD}
\end{align}
with $\boldsymbol{M}^{(s)}$ being the $N \times N$ reliability matrix at iteration $s$, $\kappa_i \in \left\{1,..., N_S \right\} \ \forall i \in \mathcal{G}_s$, $\cup_{i} \kappa_i = \left\{1,...,N_S \right\}$ and $\kappa_i < \kappa_j$ for all $i<j$. Finally, we obtain the reliability matrix $\boldsymbol{M}$ of the full set of genes $\mathcal{G}$ by normalizing each entry $ \left[ \boldsymbol{M}^{(S)} \right]_{i,j} \ i,j \in \mathcal{G}$ by $n_{i,j}$ that is the frequency of how often detecting an edge between genes $i$ and $j$ has been considered.\\
\begin{table}[h!]
\begin{framed}

{\bf{Initialization:}} $\boldsymbol{M}^{(0)} = \boldsymbol{0}_{N \times N}$; $\boldsymbol{M}_{s=0} = \boldsymbol{0}_{N_S \times N_S}$; frequency counter $n^{(0)}_{i,j} = 0$
\\
\\
{\bf{Repeat:}}
\begin{enumerate}
\item[$1:$] Select subset $\mathcal{G}_s$ of size $N_S$ from $\mathcal{G}$; draw each gene from $\mathcal{G}$ with equal probability without replacement
\item[$2:$] Update: $n^{(s+1)}_{i,j} = n^{(s)}_{i,j} + 1$ for all $i,j \in \mathcal{G}_s$ 
\item[$3:$] Estimate the DAG topology $\mathcal{D}_s$ of set $\mathcal{G}_s$ using GENIE, GI-GENIE, respectively; $\Longrightarrow$ $\boldsymbol{M}_s$    
\item[$4:$] Update reliability matrix $\boldsymbol{M}^{(s)}$ according to Eq.~(\ref{superAD}) 
\item[$7:$] Update iteration number: $s \leftarrow s+1$  
\end{enumerate}

{\bf{Until:}} $s = S$; \\
Set $ \left[ \boldsymbol{M} \right]_{i,j} = \left[ \boldsymbol{M}^{(S)} \right]_{i,j} / n^{(S)}_{i,j} \, \forall i,j \in \mathcal{G}$  

\end{framed}
\label{tab:seq_sca}
\caption{Summary of the proposed SEQSCA-algorithm}
\end{table}

In Tab.~III, we have summarized the SEQSCA-technique.
Finally, by means of the SEQSCA-technique, we are able to provide statistically stressable statements about the interactions among the genes from a large set $\mathcal{G}$ by using the GENIE or GI-GENIE algorithm, respectively, in a sequential fashion.

\section{Simulation Results}
In this section, we first demonstrate the performance of the GENIE-algorithm and the GI-GENIE-algorithm with respect to conventional techniques for simulated data and further provide 
statistically stressable statements on the interactions among the genes from a large set of genes based on real data using the proposed SEQSCA-technique. 

\subsection{Synthetic Data Results}
We have generated the ideal SK phenotypes $R(i) \in \mathbb{R}$ for all $i \in \mathcal{G}$ as well as the ideal DK phenotypes $R(i,j) \in \mathbb{R}$ for all $i,j \in \mathcal{G}:$$j>i$ according to the model of \cite{Battle} as displayed in Fig.~2, where we have corrupted the ideal SK and DK phenotypes by independently and identically distributed zero-mean Gaussian noise with variance $\sigma^2$.
Furthermore, the GI-profile similarity data $\rho(i,j) \forall i,j \in \mathcal{G}: j>i$ has been generated on the basis of the SK and DK phenotypes. 
We compare both the GENIE-algorithm and the GI-GENIE-algorithm with the well known GI-profile similarity approach (\cite{GI_corr},\cite{Battle}), where the Pearson correlation between the GI-profiles of genes $i$ and $j$ is computed and an edge in the DAG is detected if the Pearson correlation is above a pre-defined threshold $t_{\text{corr}}$, where the directionality is inferred from the selection variable $\alpha_k(i,j)$ corresponding to the least mismatch model $\mu_k(i,j)$.  
Furthermore, we compare our proposed methods with the solution of program $\text{O}_{\text{GENIE}}$ without considering set $\mathcal{L}$ as a constraint, which means simply classifying each pair $i,j$ to the least mismatch scoring hierarchical relationship class based on the SK and DK phenotypes $R(i)$ and $R(i,j)$ respectively, without ensuring that the resulting pattern of hierarchical relationship classes represents a valid DAG. 
\\
For the simulations, we consider a total of 10 genes in order to limit the Monte-Carlo simulation time.
In Fig.~7 we display the false detection percentage of edges $P_{\text{ed}}$ in the detected DAG normalized to the true number of edges $\left| \mathcal{E}_\mathcal{D} \right|$ as defined in Eq.~(\ref{misdetection}) versus the SNR.
\begin{align}
P_{\text{ed}} = \frac{ \left|   \Big( \mathcal{E}_\mathcal{D} \bigcup \hat{\mathcal{E}}_\mathcal{D}   \Big) \setminus  \mathcal{E}_\mathcal{D}  \right|  }{\left| \mathcal{E}_\mathcal{D} \right|}  \label{misdetection} 
\end{align}
In Fig.~8 we display the percentage of undetected edges $P_{\text{mis}}$ in the detected DAG normalized to the true number of edges $\left| \mathcal{E}_\mathcal{D} \right|$ as defined in Eq.~(\ref{missingedges}) versus the SNR, i.e.,
\begin{align}
P_{\text{mis}} = \frac{ \left|   \Big( \mathcal{E}_\mathcal{D} \bigcup \hat{\mathcal{E}}_\mathcal{D}   \Big) \setminus  \hat{\mathcal{E}}_\mathcal{D} \right|}{\left| \mathcal{E}_\mathcal{D} \right|}  \label{missingedges} 
\end{align}  

In Fig.~7 we observe that in the low SNR regime, the Pearson correlation based method performs best in terms of false detection percentage of edges $P_{\text{ed}}$, however it fails to improve performance with increasing SNR, because for correct directionality information of the edges this approach relies on the hierarchical relationship classes detected by method $\text{O}_{\text{GENIE}}$ without considering $\mathcal{L}$. Especially in the high SNR regime, the proposed GENIE and GI-GENIE methods clearly outperform program $\text{O}_{\text{GENIE}}$ without the topology rule-set $\mathcal{L}$ and approach and respectively reach the performance of the Pearson correlation method. 
However, the very good performance of the Pearson correlation method in terms of false detection percentage of edges $P_{\text{ed}}$ according to Eq.~(\ref{misdetection}) comes at the cost of a rather poor performance in terms of the percentage of undetected edges $P_{\text{mis}}$ according to Eq.~(\ref{missingedges}) as can be seen in Fig.~8. In particular, in terms of the percentage of undetected edges $P_{\text{mis}}$ all of the proposed methods outperform the Pearson correlation method. Note that in the high SNR regime, the GI-GENIE of combining SK, DK and GI-profile similarity data yields the best of both worlds, i.e. it shows an equivalent performance as the Pearson correlation method in terms of false detection percentage of edges $P_{\text{ed}}$, as well as an improvement of the strong performance of the GENIE method in terms of the percentage of undetected edges $P_{\text{mis}}$.
  
\begin{figure}[htb]
  \centering
  \def\svgwidth{\columnwidth}
  \definecolor{mycolor1}{rgb}{0.00000,1.00000,1.00000}%
\begin{tikzpicture}

\begin{axis}[%
width=0.38\textwidth,
height=0.4\textwidth,
at={(0\textwidth,0\textwidth)},
scale only axis,
separate axis lines,
every outer x axis line/.append style={black},
every x tick label/.append style={font=\color{black}},
xmin=0,
xmax=40,
xlabel={SNR in [db]},
xmajorgrids,
every outer y axis line/.append style={black},
every y tick label/.append style={font=\color{black}},
ymin=0,
ymax=1.4,
ylabel={False detection percentage $P_\text{ed}$ according to Eq.~(\ref{misdetection})},
ymajorgrids,
axis background/.style={fill=white},
legend style={legend cell align=left,align=left,draw=black}
]
\addplot [color=red,solid,line width=2.0pt,mark=square,mark options={solid}]
  table[row sep=crcr]{%
0	0.254532677590727\\
6.66666666666667	0.264124102613073\\
13.3333333333333	0.244980301754489\\
20	0.185882392396905\\
26.6666666666667	0.120078725196372\\
33.3333333333333	0.0552427217390452\\
40	0.0244664306281953\\
};
\addlegendentry{\scriptsize{Pearson correlation}};

\addplot [color=green,solid,line width=2.0pt,mark=square,mark options={solid}]
  table[row sep=crcr]{%
0	1.10990108178285\\
6.66666666666667	1.20414925148381\\
13.3333333333333	1.1806203842597\\
20	0.949378418902219\\
26.6666666666667	0.699893491312609\\
33.3333333333333	0.443853087272205\\
40	0.315181399547305\\
};
\addlegendentry{\scriptsize{$\text{O}_\text{GENIE} \ \text{without} \ \mathcal{L}$} };

\addplot [color=blue,dotted,line width=2.0pt,mark=+,mark options={solid}]
  table[row sep=crcr]{%
0	1.10657693413827\\
6.66666666666667	1.09269472684179\\
13.3333333333333	0.960584921657289\\
20	0.708030059852428\\
26.6666666666667	0.360657664639282\\
33.3333333333333	0.175969685135126\\
40	0.121998903766512\\
};
\addlegendentry{\scriptsize{$\text{O}_\text{GENIE}$}  };

\addplot [color=mycolor1,dotted,line width=2.0pt,mark=+,mark options={solid}]
  table[row sep=crcr]{%
0	1.10136275197111\\
6.66666666666667	1.08355038425627\\
13.3333333333333	0.876915698207881\\
20	0.475033419753621\\
26.6666666666667	0.12260758604141\\
33.3333333333333	0.0354810112273347\\
40	0.0260275361893009\\
};
\addlegendentry{\scriptsize{$\text{O}_\text{GI-GENIE}$}  };

\end{axis}
\end{tikzpicture}%
  \caption{$D_{\text{ed}}$ versus SNR; $t_{\text{corr}}=.6$; 200 Monte Carlo runs; $\lambda_d=1$,
	$\lambda_s = 8e-6$, $\lambda_c = 20$, $\lambda_p = 16$}
\end{figure}
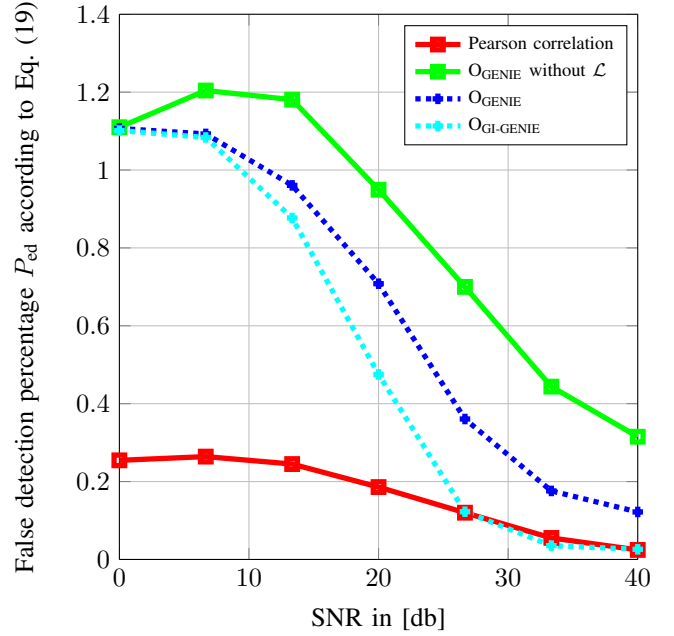

\begin{figure}[htb]
  \centering
  \def\svgwidth{\columnwidth}
  \definecolor{mycolor1}{rgb}{0.00000,1.00000,1.00000}%
\begin{tikzpicture}

\begin{axis}[%
width=0.38\textwidth,
height=0.4\textwidth,
at={(0\textwidth,0\textwidth)},
scale only axis,
separate axis lines,
every outer x axis line/.append style={black},
every x tick label/.append style={font=\color{black}},
xmin=0,
xmax=40,
xlabel={SNR in [db]},
xmajorgrids,
every outer y axis line/.append style={black},
every y tick label/.append style={font=\color{black}},
ymin=0,
ymax=1,
ylabel={Percentage of undetected edges $P_\text{mis}$ according to Eq.~(\ref{missingedges})},
ymajorgrids,
axis background/.style={fill=white},
legend style={legend cell align=left,align=left,draw=black}
]
\addplot [color=red,solid,line width=2.0pt,mark=square,mark options={solid}]
  table[row sep=crcr]{%
0	0.936439748074423\\
6.66666666666667	0.879427872372726\\
13.3333333333333	0.799009699472354\\
20	0.670442553189845\\
26.6666666666667	0.505909722630311\\
33.3333333333333	0.333419753041077\\
40	0.272442332564237\\
};
\addlegendentry{\scriptsize{Pearson correlation} };

\addplot [color=green,solid,line width=2.0pt,mark=square,mark options={solid}]
  table[row sep=crcr]{%
0	0.85113233429797\\
6.66666666666667	0.780072159458189\\
13.3333333333333	0.705459741174718\\
20	0.583250826508373\\
26.6666666666667	0.392602817525612\\
33.3333333333333	0.201148436122701\\
40	0.0936542965548383\\
};
\addlegendentry{\scriptsize{$\text{O}_\text{GENIE} \ \text{without} \ \mathcal{L}$}};

\addplot [color=blue,dotted,line width=2.0pt,mark=+,mark options={solid}]
  table[row sep=crcr]{%
0	0.830845188952775\\
6.66666666666667	0.743075127894981\\
13.3333333333333	0.60854371183512\\
20	0.469863062032567\\
26.6666666666667	0.229514522242464\\
33.3333333333333	0.0940153682102211\\
40	0.0492837563623167\\
};
\addlegendentry{\scriptsize{$\text{O}_\text{GENIE}$}};

\addplot [color=mycolor1,dotted,line width=2.0pt,mark=+,mark options={solid}]
  table[row sep=crcr]{%
0	0.825382439284143\\
6.66666666666667	0.737355321965616\\
13.3333333333333	0.564603213607858\\
20	0.316232450212326\\
26.6666666666667	0.081522713805802\\
33.3333333333333	0.0294258829405888\\
40	0.0234946758607749\\
};
\addlegendentry{\scriptsize{$\text{O}_\text{GI-GENIE}$}};

\end{axis}
\end{tikzpicture}%
  \caption{$D_{\text{mis}}$ versus SNR; $t_{\text{corr}}=.6$; 200 Monte Carlo runs; $\lambda_d=1$,
	$\lambda_s = 8e-6$, $\lambda_c = 20$, $\lambda_p = 16$}
\end{figure}
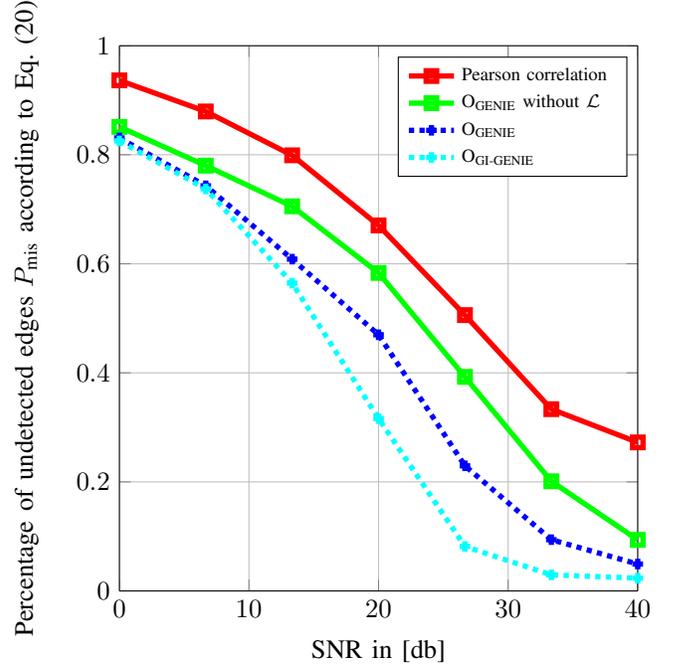

\subsection{Real Data Results}
In this section, we apply the above mentioned SEQSCA-algorithm to the dataset of \cite{drygin}, in order to yield statistically stressable statements about the interactions among the genes considered in \cite{drygin}.   
In \cite{drygin}, the organism under study is yeast and the cell process measured is the colony size as a proxy of fitness.
For the sake of computation time, we restrict the full set of genes under study in \cite{drygin}, i.e., $\mathcal{G}$, to contain the first 200 genes of the \textit{query list} of \cite{drygin}. Furthermore, the length of the sequence of subsets $S$ is set to 50000 and the subset size $N_S$ of each selected set $\mathcal{G}_s$ in the SEQSCA-algorithm is set to 10.
\begin{figure}[h]
  \centering
  \def\svgwidth{\columnwidth}
  \begin{tikzpicture}

\begin{axis}[%
width=0.32\textwidth,
height=0.4\textwidth,
at={(0\textwidth,0\textwidth)},
scale only axis,
point meta min=0,
point meta max=1,
axis on top,
separate axis lines,
every outer x axis line/.append style={black},
every x tick label/.append style={font=\color{black}},
xmin=0.5,
xmax=201.5,
xlabel={Gene index},
every outer y axis line/.append style={black},
every y tick label/.append style={font=\color{black}},
y dir=reverse,
ymin=0.5,
ymax=201.5,
ylabel={Gene index},
axis background/.style={fill=white},
title={},
colormap/jet,
colorbar,
colorbar style={separate axis lines,every outer x axis line/.append style={black},every x tick label/.append style={font=\color{black}},every outer y axis line/.append style={black},every y tick label/.append style={font=\color{black}}}
]
\addplot [forget plot] graphics [xmin=0.5,xmax=201.5,ymin=0.5,ymax=201.5] {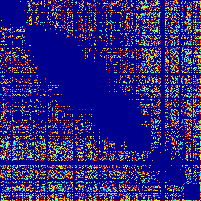};
\end{axis}
\end{tikzpicture}%
  \caption{Reliability matrix $\boldsymbol{M}$ for the SEQSCA-technique using GENIE-algorithm; $S=50000$ subsets considered; subset size $N_S = 10$}
	\label{fig:superAD_GENIE}
\end{figure}
\\
In Fig.~\ref{fig:superAD_GENIE} we show the reliability matrix $\boldsymbol{M} \in \left[ 0,1\right]^{200 \times 200}$ for $S=50000$ subsets, where we have applied the SEQSCA-technique with the GENIE-algorithm for estimating in each iteration $s$ the topology of DAG $\mathcal{D}_s$ underlying the data of subset $\mathcal{G}_s$. 
It can be observed that only for a few pairs of genes $i,j \in \mathcal{G}$ edges have been detected with high empirical probability, which is is in line with the common assumption in genomics research that genetic interactions are generally sparse. 
Furthermore, in Fig.~\ref{fig:superAD_GENIE} the empirical probability of an edge to be detected is either smaller than $30\%$ or bigger than $70\%$ for $87\%$ of the gene pairs $i,j \in \mathcal{G}$. Hence, for most gene pairs the proposed SEQSCA-technique, using the GENIE-algorithm, is able to provide a statistically stressable statement whether a two genes $i,j$ interact with each other or not.    
\begin{figure}[h]
  \centering
  \def\svgwidth{\columnwidth}
  \begin{tikzpicture}

\begin{axis}[%
width=0.32\textwidth,
height=0.4\textwidth,
at={(0\textwidth,0\textwidth)},
scale only axis,
point meta min=0,
point meta max=1,
axis on top,
separate axis lines,
every outer x axis line/.append style={black},
every x tick label/.append style={font=\color{black}},
xmin=0.5,
xmax=201.5,
xlabel={Gene index},
every outer y axis line/.append style={black},
every y tick label/.append style={font=\color{black}},
y dir=reverse,
ymin=0.5,
ymax=201.5,
ylabel={Gene index},
axis background/.style={fill=white},
title={},
colormap/jet,
colorbar,
colorbar style={separate axis lines,every outer x axis line/.append style={black},every x tick label/.append style={font=\color{black}},every outer y axis line/.append style={black},every y tick label/.append style={font=\color{black}}}
]
\addplot [forget plot] graphics [xmin=0.5,xmax=201.5,ymin=0.5,ymax=201.5] {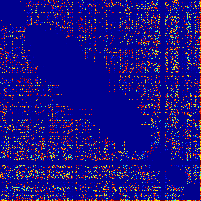};
\end{axis}
\end{tikzpicture}%
  \caption{Reliability matrix $\boldsymbol{M}$ for the SEQSCA-technique using GI-GENIE-algorithm; $S=50000$ subsets considered; subset size $N_S = 10$; $\lambda_d=1e3$,
	$\lambda_s = 8e-6$, $\lambda_c = 1$, $\lambda_p = .85$}
	\label{fig:superAD_GI_GENIE}
\end{figure}
In Fig.~\ref{fig:superAD_GENIE} we show the reliability matrix $\boldsymbol{M} \in \left[ 0,1\right]^{200 \times 200}$ for again $S=50000$ subsets, where we have applied the SEQSCA-technique with the GI-GENIE-algorithm for estimating in each iteration $s$ the topology of DAG $\mathcal{D}_s$ underlying the data of subset $\mathcal{G}_s$. 
In this case, the GI-profile data $\rho(i,j) \ \forall i,j \in \mathcal{G}: i\neq j$ has been computed based on the SK and DK phenotypes of \cite{drygin} according to Eq.~(\ref{GI_data}).
In Fig.~\ref{fig:superAD_GENIE}, the empirical probability that two genes interact with each other, is either smaller than $30\%$ or bigger than $70\%$ for $95\%$ of the gene pairs $i,j \in \mathcal{G}$. Thus, the incorporation of GI-profile data into the topology estimation process in each iteration $s$ of the SEQSCA-technique enhances the statistical reliability of the statements made by the SEQSCA-technique, i.e., whether two genes $i,j$ interact directly interact with each other or not.

\section{Conclusion}
In this paper we have considered the problem of learning the interactions between genes in a genetic network. 
Our approach for computing the interactions among the genes, i.e., the genetic-interaction map which is essentially a DAG, uses the interaction models of \cite{Battle} and combines it with GI-profile data, that can be computed based on the Pearson correlation of the SK and DK phenotypes of the set of genes under study, or it can be extracted from some data base where other types of GI-profile data are stored. 
Furthermore, we have proposed an scalability approach, called SEQSCA, which uses the proposed GENIE and GI-GENIE-methods to yield statistically reliable statements, whether two genes out of large set of genes interact with each other or not.

%
%
%
%


%

\ifCLASSOPTIONcaptionsoff
  \newpage
\fi


\begin{thebibliography}{1}

\bibitem{IEEEhowto:Snijder}
B.~Snijder, P.~Liberali, M.~Frechin, T.~Stoeger and L.~Pelkmans, \emph{Predicting functional gene interactions with the hierarchical interaction score},\hskip 1em plus
  0.5em minus 0.4em\relax Nature Methods, VOL.10 NO.11, November 2013
	
\bibitem{IEEEhowto:Baryshinkova}
A.~Baryshinkova et al.  \emph{Quantitative analysis of fitness and genetic interactions in yeast on a genome scale},\hskip 1em plus
  0.5em minus 0.4em\relax Nature Methods, December 2010	
	
			\bibitem{Collins}
S.R.~Collins, A.~Roguev, N.J.~Krogan \emph{Quantitative Genetic Interaction Mapping Using the E-MAP
Approach},\hskip 1em plus
  0.5em minus 0.4em\relax Methods Enzymol., 2010	
	
			\bibitem{Linden}
R.O.~Linden, V.P.~Eronen, T.~Aittokallio \emph{Quantitative maps of genetic interactions in yeast - Comparative evaluation and integrative analysis},\hskip 1em plus
  0.5em minus 0.4em\relax BMC Systems Biology, 2011	
	
				\bibitem{Dixon}
S.J.~Dixon, M.~Constanzo, A.~Baryshinkova, B.~Andrews, C.~Boone \emph{Systematic Mapping of Genetic Interaction Networks},\hskip 1em plus
  0.5em minus 0.4em\relax The Annual Review of Genetics, 2009	
	
	
	\bibitem{Brock}
G.N.~Brock et al. \emph{Methods for detecting gene × gene interaction in multiplex 
extended pedigrees},\hskip 1em plus
  0.5em minus 0.4em\relax BMC Genetics, 2005
	
	
	\bibitem{Battle}
A.~Battle, M.C.~Jonikas, P.~Walter, J.S.~Weissman and D.~Koller \emph{Automated identification of pathways from quantitative genetic interaction data},\hskip 1em plus
  0.5em minus 0.4em\relax Molecular Systems Biology 6, 2010	
		
	
	\bibitem{LIP1}
T.C.~Hu, A.B.~Kahng \emph{Linear and Integer Programming in Practice},\hskip 1em plus
  0.5em minus 0.4em\relax, Springer International Publishing, Schweiz, 2016, ISBN-10: 3319239996
	
\bibitem{LIP2}
G.~Sierksma \emph{Linear and Integer Programming: Theory and Practice, Second Edition},\hskip 1em plus
  0.5em minus 0.4em\relax, CRC Press, Boca Raton FL 33487-2741, 2001, ISBN-10: 0824706730
	
\bibitem{LIP3}
G.~Sierksma, Y.~Zwols \emph{Linear and Integer Optimization: Theory and Practice, Third Edition},\hskip 1em plus
  0.5em minus 0.4em\relax, CRC Press, Boca Raton FL 33487-2741, 2015, ISBN-10: 1498710166
	
\bibitem{LIP4}
E.~Demirel, N.~Demirel, H.~G\"okcen \emph{A mixed integer linear programming model to optimize reverse logistics activities of end-of-life vehicles in Turkey},\hskip 1em plus
  0.5em minus 0.4em\relax, Journal of Cleaner Production, 2016

\bibitem{LIP5}
C.H.~Antunes, M.J.~Alves, J.~Climaco\emph{Multiobjective Linear and Integer Programming},\hskip 1em plus
  0.5em minus 0.4em\relax, Springer International Publishing, Schweiz, 2016, ISBN-13: 9783319287447
	
\bibitem{LIP6}
M.~Diaby, M.H.~Karwan\emph{Advances in Combinatorial Optimization},\hskip 1em plus
  0.5em minus 0.4em\relax, World Scientific Publishing Co. Pte. Ltd., Singapore, 2016, ISBN-10: 9814704873
	
	

	\bibitem{SGA}
A.H.Y.~Tong et al. \emph{Systematic genetic analysis with ordered arrays of yeast deletion mutants},\hskip 1em plus
  0.5em minus 0.4em\relax Science 294, 2001	
	
\bibitem{Graph}
R.~Diestel \emph{Graphentheorie},\hskip 1em plus
  0.5em minus 0.4em\relax Springer-Verlag, Heidelberg,2012, ISBN 978-3-642-14911-5	
	
\bibitem{IEEEhowto:Papa}
C.H.~Papadimitriou, K.~Steiglitz \emph{Combinatorial optimization: algorithms and complexity},\hskip 1em plus
  0.5em minus 0.4em\relax Mineola NY, 1998, ISBN 0486402584

\bibitem{GI_map1}
A.~Jaimovich et al. \emph{Modularity and directionality in genetic interaction maps},\hskip 1em plus
  0.5em minus 0.4em\relax Nature Methods, 2010, Vol. 26 
	
\bibitem{GI_map2}
A.~Baryshinkova, M.~Constanzo, C.L.~Myers, B.~Andrews, C.~Boone  \emph{Genetic Interaction Networks: Toward an Understanding of Heritability},\hskip 1em plus
  0.5em minus 0.4em\relax Annual Review of Genomics and Human Genetics, 2013
	
	\bibitem{GI_map4}
A.~Rogueav et al. \emph{Quantitative genetic-interaction mapping in mammalian cells},\hskip 1em plus
  0.5em minus 0.4em\relax Nature Methods, February 2013, Vol. 10
	
		\bibitem{GI_map5}
M.~Constanzo et al. \emph{The genetic landscape of a cell},\hskip 1em plus
  0.5em minus 0.4em\relax Science 327:425-431, 2010
	
		\bibitem{BB1}
V.~Balakrishnan, S.~Boyd, S.~Balemi \emph{Branch and bound algorithm for computing the minimum stability degree of parameter-dependent linear systems},\hskip 1em plus
  0.5em minus 0.4em\relax International Journal of Robust and Nonlinear Control,1(4):295–317, October–December 1991
	
	
\bibitem{BB2}
E.L.~Lawler, D.E.~Wood \emph{Branch-and-bound methods: A survey},\hskip 1em plus
  0.5em minus 0.4em\relax Operations Research, 14:699–719, 1966
	
\bibitem{BB3}
R.E.~Moore \emph{Global optimization to prescribed accuracy},\hskip 1em plus
  0.5em minus 0.4em\relax Computers and Mathematics with Applications, 21(6/7):25–39, 1991
	
\bibitem{BB4}
Y.~Cheng, M.~Pesavento \emph{Joint Rate Adaptation and Downlink Beamforming Using Mixed Integer Conic Programming},\hskip 1em plus
  0.5em minus 0.4em\relax IEEE Transactions on Signal Processing, 2013
		
\bibitem{BB5}
Y.~Cheng, M.~Pesavento \emph{An Optimal Iterative Algorithm for Codebook-Based Downlink Beamforming},\hskip 1em plus
  0.5em minus 0.4em\relax IEEE Signal Processing Letters, 2013	
	
\bibitem{BB6}
Y.~Cheng, M.~Pesavento \emph{Joint Optimization of Source Power Allocation and Distributed Relay Beamforming in Multiuser Peer-to-Peer Relay Networks},\hskip 1em plus
  0.5em minus 0.4em\relax IEEE Transactions on Signal Processing, Vol.~60, No~6, pp. 2395-2404, 2012	
		
\bibitem{BB7}
Y.~Cheng, M.~Pesavento, A.~Philipp \emph{Joint Network Optimization and Downlink Beamforming for CoMP Transmissions using Mixed Integer Conic Programming},\hskip 1em plus
  0.5em minus 0.4em\relax  IEEE Transactions on Signal Processing, Vol.~61, 2013			

\bibitem{GI_corr}
M.~Babu et al. \emph{Quantitative Genome-Wide Genetic Interaction Screens
Reveal Global Epistatic Relationships of Protein
Complexes in Escherichia coli},\hskip 1em plus
  0.5em minus 0.4em\relax PLOS Genetics, February 2014, Volume 10
	
	\bibitem{drygin}
M.~Costanzo et al. \emph{DRYGIN - Data Repository of Yeast Genetic Interactions},\hskip 1em plus
  0.5em minus 0.4em\relax Terence Donnelly Centre for Cellular and Biochemical Research, University of Toronto,
	http://drygin.ccbr.utoronto.ca/~costanzo2009/
	
		\bibitem{dag_paper}
A.~Shojaie, G.~Michailidis \emph{Discovering graphical Granger causality using the truncating
lasso penalty},\hskip 1em plus
  0.5em minus 0.4em\relax Department of Statistics, University of Michigan, ECCB 2010, Vol. 26
	
\bibitem{fabio_eusipco}
F.~Nikolay, M.~Pesavento \emph{Learning Directed-Acyclic-Graphs from Large-Scale Double-Knockout Experiments},\hskip 1em plus
  0.5em minus 0.4em\relax C, Communications System Group, TU Darmstadt, EUSIPCO 2016, Budapest
	
		\bibitem{supp_mat}
		Supplementary Material: \emph{\url{https://www2.spg.tu-darmstadt.de/~fnikolay/supp_journal.pdf}  }
	
\end{thebibliography}
\end{document}